\documentclass[english]{solarphysics}
\usepackage{graphicx}
\usepackage[optionalrh,solaromanenum]{spr-sola-addons} 
\usepackage{url}                         
\usepackage[pdfborder={0 0 0},colorlinks=true,urlcolor=blue,linkcolor=blue,citecolor=blue, breaklinks]{hyperref} 
\include{sola_labels}

\makeatletter

\newcommand{\etal}{  {\it et al.}}

\newcommand{\noun}[1]{\textsc{#1}}

\newcommand{\comment}[2]{#2}

\makeatother

\usepackage{babel}

\begin{document}

\begin{article}

\begin{opening}

\title{Polarimetric Properties of Flux-Ropes and Sheared Arcades in Coronal Prominence Cavities}

%
\author{L.A.~\surname{Rachmeler}$^{1,2}$\sep
        S.E.~\surname{Gibson}$^{3}$\sep
        J.B.~\surname{Dove}$^{4}$\sep
        C.R.~\surname{DeVore}$^{5}$\sep
        Y.~\surname{Fan}$^{3}$
       }

\runningauthor{L.A. Rachmeler \etal}
\runningtitle{Polarimetric Properties of Theoretical Coronal Prominence Cavities}

%
  \institute{$^{1}$  School of Mathematics and Statistics, University of St. Andrews, North Haugh, St. Andrews, Fife, KY16 9SS, UK \\
             $^{2}$ Royal Observatory of Belgium, Avenue Circulaire 3, 1180 Brussels, Belgium\\ 
  email: \url{rachmeler@oma.be} \\ 
              $^{3}$ High Altitude Observatory, NCAR, P.O. Box 3000, Boulder, CO, 80307-3000, USA\\ 
             email: \url{sgibson@ucar.edu} \ \\
             $^{4}$ Physics Department, Metropolitan State University of Denver, NC 3123, Campus Box 69, 1201 5th Street, Denver, CO 80204-2005, USA\\
             $^{5}$ Laboratory for Computational Physics and Fluid Dynamics, Naval Research Laboratory, Washington, DC 20375, USA
             }


\begin{abstract}
The coronal magnetic field is the primary driver of solar dynamic
events. Linear and circular polarization signals of certain infrared
coronal emission lines contain information about the magnetic field,
and to access this information, either a forward or an inversion method
must be used. We study three coronal magnetic configurations that
are applicable to polar-crown filament cavities by doing forward
calculations to produce synthetic polarization data. We analyze these
forward data to determine the distinguishing characteristics of each
model. We conclude that it is possible to distinguish between cylindrical
flux ropes, spheromak flux ropes, and sheared arcades using coronal
polarization measurements. If one of these models is found to be consistent
with observational measurements, it will mean positive identification
of the magnetic morphology that surrounds certain quiescent filaments,
which will lead to a greater understanding of how they form and why
they erupt. 
\end{abstract}

%

\end{opening}

%
\section{Introduction}

In order to understand coronal evolution and be able to predict dynamic
events such as coronal mass ejections (CMEs) and solar flares, we
need to measure the coronal magnetic field. However, making this
measurement is not a trivial task. Much of this difficulty is due
to the optically thin nature of the coronal plasma, and the relatively
weak intensities of coronal emission lines. Measuring the field will
bring resolution to many long-standing debates about the corona, including
the magnetic-field morphology surrounding prominences. 

Prominences can be extremely stable on the solar surface; some polar
crown filaments survive for many rotations \cite{Gibson2006_cavities},
but they are also known to erupt suddenly. When prominences are seen
on the limb, and are aligned with an observers' line-of-sight
(LOS), they are often seen to be embedded in coronal cavities. These
cavities are typically depleted in density by a factor of about two
relative to the surrounding streamer \cite{Fuller2009,Schmit2011}. Cavities
are the coronal manifestation of the magnetic system that also includes
the filament channel and the prominence \cite{Hudson1999,Gibson2006_cavities,Heinzel2008,gibson2010-cavities}.
Because the cavity comprises the bulk of the system volume, the magnetic
structure of the system can be determined from measurements of the
cavity. The cavity--prominence structure is known to erupt bodily as
a CME \cite{Maricic2004,Regnier2011}. The initiation of these eruptions
depends critically on the magnetic field threading through, and around,
the prominence.

Flux-rope models, with magnetic field wrapped around a distinct axis,
and sheared-arcade models, without such an axis, have both been posited
as possible morphologies of prominences and their surrounding magnetic
field (see \opencite{Mackay2010} and references therein). 
Flux-rope systems have also been explicitly compared
to cavities \cite{LowHund1995,Gibson2006_cavities,Dove2011}.
Both morphological models contain dipped field lines where mass can
cool and condense into prominence material. Coronal polarization measurements
could provide a means of distinguishing between these structures,
which would lead to a greater understanding of not only the quiescent
nature of prominences and cavities, but how they are formed, and how
they destabilize and erupt. Our objective in this article is to determine the 
characteristic polarization signatures of these different models of prominence cavities.

In Section~\ref{sec:measurement} we discuss the difficulties of measuring the 
coronal magnetic field and the specifics of the Stokes vector in the Fe \noun{xiii} 
1074.7~nm coronal emission line. 
Section~\ref{sec:forward model} describes the forward 
calculations, followed by details of the three individual coronal models 
in Section~\ref{sec:mhd}.
We analyze these synthetic forward-modeled
observations and look for distinguishing features, which can be compared
to solar observations of prominence--cavity systems to identify
magnetic morphologies.  We present these features in Section~\ref{sec:Results} and conclude with a discussion of our results in Section~\ref{sec:discussion}.

\section{Measuring the Coronal Magnetic Field  \label{sec:measurement} }

There are several
methods currently employed to determine the magnetic field
in the corona. Given that the thermal conductivity along the field is high 
compared to across the field, the \emph{bright
loops} seen in extreme ultra-violet (EUV) and X-ray images of the
corona trace out magnetic field lines. Using EUV images, it is possible
to follow these lines subject to projection effects \cite{Aschwanden1999}.
However, this method does not produce a measure of the magnitude of the field,
and full three-dimensional (3D) traces require tomographic inversions
and/or stereoscopic methods which can be tricky in the optically thin coronal
plasma. Also, this technique only provides magnetic morphology information
on specific bright loops, and not the full volume containing
the magnetic field. 

Measuring \emph{Faraday rotation} along a LOS to a known radio source
supplies information about the LOS magnetic field, if the plasma density
is known. This technique has been used to study the corona \cite{Patzold1987,Jensen_phd2007},
but a limited number of sight lines exist along which this technique
is valid. \emph{Gyroresonant emission} in radio wavelengths is related
to the total magnetic field strength in the emitting region, but is
limited to areas of strong magnetic field ($>200$ Gauss) such as active
regions \cite{White1997}, although instrumentation with a broader
spectral range, such as the {\it Frequency Agile Solar Radiotelescope} (FASR: \opencite{Bastian2004}),
would allow for more extensive application. Observation of modified
\emph{bremsstrahlung emission} in radio and microwave frequencies
provides information on the LOS magnetic field in the corona on the
disk. These measurements sample a thin layer in the lower corona/upper
chromosphere, not in the full coronal volume \cite{Gelfreikh1994,Grebinkij2000}.
In some coronal emission lines, particularly in forbidden magnetic-dipole emission lines in the infrared, \emph{resonant scattering}
of anisotropic light in a magnetized plasma can produce polarized
emission subject to the Hanle and the Zeeman effects. We will now discuss
the use of coronal emission-line polarization in more detail and further
proceed to forward modeling of this emission.

\subsection{Coronal Stokes Vector\label{sub:Coronal-Stokes-vector}}

\inlinecite{Charvin1965} was one of the first to show that linear-polarization
signals from forbidden coronal emission-line transitions could be
used to determine the plane-of-sky (POS) magnetic-field direction.
\inlinecite{Harvey1969} was the first to attempt to use circular polarization
to measure the LOS magnetic-field strength. Compared to modern technology,
early detectors of the coronal Stokes vector had significantly lower signal-to-noise
ratios, coarser resolution, and required longer integration times,
which in turn limited temporal resolution. One such early detector was the {\it Coronal Emission Line Polarimeter} \cite{Querfeld1977},
which was a scanning photodiode polarimeter. A full-Sun measurement
with this instrument would typically take about two hours and contain
1408 data points from 1.01~R$_{\odot}$ to 1.65~R$_{\odot}$ \cite{Querfeld1977,Arnaud1987}.

In the intervening $40$ years, 
there has been steady progress in the field of coronal polarization.
Today, there are two main coronal polarimeters currently in use: 
The first is the {\it Optical Fiber-bundle Imaging
Spectropolarimeter} (OFIS) on the {\it Solar Observatory for Limb Active
Regions and Coronae} (SOLARC) at Mt. Haleakala \cite{Lin2004}. It
generates $128$ spectra from a $16\times8$ fiber optic array that
subtends $5\times2.5$ arc minutes. It is capable of measuring full
Stokes profiles at each of the $128$ positions. This instrument has
been used to successfully measure the linear-polarization strength
and direction in the Fe\noun{ xiii} 1074.7 nm line, and to determine
a LOS field strength from circular-polarization signals above an active
region \cite{Lin2004,Liu2008}. The second instrument is the {\it Coronal
Multi-channel Polarimeter} (CoMP: \opencite{Tomczyk2008}), which is installed at
the Mauna Loa Solar Observatory, and began taking full-corona measurements
in October 2010. CoMP is an imaging coronagraph polarimeter with a
tunable birefringent filter capable of detecting the Fe \noun{xiii}
1074.7 nm and 1079.8 nm lines as well as the He \noun{i} 1083 nm line.
The new CoMP observations provide, for the first time, daily full-Sun
observations of the magnetic field in the corona. The primary observables
of CoMP are the four Stokes parameters ($I$, $Q$, $U$, $V$). 

\comment{include LOS integration somewhere here!}

These observations are taken above the solar limb in the corona, which is optically thin in these wavelengths, and thus the measurements contain information from and extended LOS source. The polarization signal strength is weak compared to the line intensity (linear
polarization/intensity $\thickapprox 10^{-2}$ and circular polarization/intensity $\thickapprox10^{-4}$
for a one-Gauss field: \opencite{Arnaud1987}; \opencite{Lin2000}). It takes on the order of a few minutes to obtain
a useable full-Sun linear-polarization measurement with CoMP, and
circular-polarization measurements of sufficient signal-to-noise are
made by averaging over an hour of data \cite{Tomczyk2008}. The polarization is the result of resonant scattering of
anisotropic incident radiation by highly ionized coronal plasma
in the presence of an external magnetic field. However, different
aspects of this unified process dominate the linear- and
circular-polarization signals of the coronal emission lines ({\it e.g.} \opencite{Casini1999};  \opencite{Casini2002}; \opencite{Judge2007}; \opencite{Rachmeler2012}, and references therein). We restrict the rest of this discussion to the Fe \noun{xiii} $1074.7$ nm coronal emission line. This is also the line used in the forward calculations. 

The linear-polarization signal is completely dominated by the Hanle effect: a depolarization of scattered light associated with a radiation-induced population imbalance of the
atomic levels \cite{Trujillo2001}. The \emph{atomic alignment} [$\sigma$] describes this population imbalance. The transverse
Zeeman effect, which is due to the energy splitting of the magnetic
sublevels by the coronal field, is a secondary source of linear polarization, because the field strength is small (the transverse Zeeman effect is quadratic in the field
strength) compared to the thermal width of these coronal lines. The
Larmor frequency is larger than the inverse lifetime of
the excited state, so the linear polarization signal occurs in the
strong-field regime, also known as the \emph{saturated} Hanle effect.
In this regime, the linear-polarization strength and direction is
dependent on the angle of the magnetic field, but no information about
the magnitude is contained in the signal. 

The strength of the \emph{total linear polarization} [$L=\sqrt{Q^{2}+U^{2}}$]
(same as $P$ in \opencite{Dove2011}; \opencite{Rachmeler2012}), is dependent
on the angle [$\Theta$] between the LOS and the local magnetic-field
vector. Specifically, $L~\propto~\sin^{2}\Theta$ such that \textbf{$L$} is strong when the magnetic field is in the POS, and weak when the
magnetic field is along the LOS. The relative strengths of
$Q$ and $U$ are used to determine the \emph{azimuth angle} [$\Psi$; $U/Q=\tan2\Psi$],
the POS angle of of the LOS integrated magnetic field.
There is a $90{^\circ}$ ambiguity known as the Van
Vleck effect \cite{vanVleck1925,House1977} such that the magnetic-field direction could be parallel or perpendicular to the measured $\Psi$. When the local magnetic
field is at the Van Vleck angle of roughly $54.7{^\circ}$ with respect
to solar radial, the light becomes unpolarized, and the strength of
$L$ goes to zero. When the magnetic field is less than $54.7{^\circ}$
from radial, $\Psi$ is parallel
to the direction of the POS component of $\mathbf{B}$, but switches
to perpendicular when that angle is surpassed (see, {\it e.g.} Figure \ref{fig:yfan_results}(c)).
The Van Vleck effect results in linear polarization directions
in the corona that are mostly radial \cite{Arnaud1987}. If the location of the Van Vleck
inversion can be identified, the $90{^\circ}$ ambiguity can be removed,
although a $180{^\circ}$ ambiguity remains.

A measure of the magnetic-field strength is not possible with linear polarization in this regime, but the \emph{circular polarization} does contain information about its magnitude along the LOS. The Stokes $V$ profile is proportional
to $B\cos\Theta$. The longitudinal Zeeman effect (which is linear in the
field strength) is the main contributor to the
circular-polarization signal. However, the atomic alignment can yield a significant correction to this signal, changing
the amplitude of the anti-symmetric $V$ profile, and therefore
affecting the diagnostics of the magnetic field strength \cite{Casini1999}.

Stokes $I$, $Q$, $U$, and $V$ are all dependent on the plasma
parameters in the emitting region. They are weakly dependent on the
temperature as long as the emission line is excited. 
All of the Stokes
components are directly weighted by the density. At a given location
along the LOS, the density dependence cancels when analyzing the \emph{relative}
polarizations [$L/I$ and $V/I$] but this is not the case in
a signal that is integrated along the LOS.  For an integrated measurement, 
the signal will be dominated by those areas along the LOS that have 
the highest density. Since collision tend to equalize the sublevel populations, a density dependence also enters in to the Stokes vector through $\sigma$.

There are two general methods for interpreting the coronal-polarization
measurements. The first is by inverting the signals into physical
properties of the magnetic field. This is the approach taken with
photospheric polarization data. However,
because the plasma is optically thin in the corona, the signal is
coming from an elongated source along a LOS. Inversions generally solve for
a single point of emission, so not all of the calculations will converge
to a solution. Inversions of these polarization signals require numerous
initial assumptions about the emitting plasma, and are quite difficult
due to multiple integrals that must be inverted. These calculations
are known to be ill-posed \cite{Judge2007}. Information about atomic
level-populations, and hence the plasma parameters, at each point
along the LOS is required to solve the POS magnetic-field direction.
In order to determine the field strength and direction everywhere,
tomographic inversions are needed. The tomographic inversion process
requires multiple viewpoints of the field, and if only one is available,
as is currently the case, solar rotation must be used to generate
these viewpoints \cite{Kramar2006,Kramar2007}. This adds the additional
assumption that the coronal field does not change appreciably over
rotational timescales. An alternate approach to extracting information
from coronal-polarization data is forward modeling. 

Our forward technique involves creating simulated polarimetric observables
from models of the corona \cite{Judge2001,Judge2006}. In the work
presented here, we use this technique to study the differences between
several pre-CME magnetic morphologies, and expand upon the work begun
by \inlinecite{Judge2006}. The ultimate goal of this research is to determine
if it is possible to use coronal polarization to positively identify
flux ropes, or other magnetic morphologies, in the cavities that surround
pre-CME filaments.

\comment{The polarization is the result of resonant scattering of an anisotropic radiation field by a
highly ionized atom in the presence of an external magnetic field. The linear- and circular-polarization signals are dominated by different physical mechanisms within this process: the Hanle Effect and the Zeeman Effect 
(see, for example, \opencite{Casini1999};  \opencite{Casini2002}; \opencite{Judge2007}; \opencite{Rachmeler2012}, and references therein). The Hanle Effect is a magnetically dependent depolarization of the scattered light. The Zeeman Effect removes the energy degeneracy of the magnetic substates due to the presence of the magnetic field. For the Fe \noun{xiii} $1074.7$ nm line, Hanle is dominant (0th and 1st order in the Zeeman splitting) in the linear polarization, and Zeeman is dominant (1st order) in the circular polarization. Thus, the circular-polarization signal strength is much weaker than that of the linear polarization. }

\section{Description of the Forward Calculations\label{sec:forward model}}

The basic procedure of our forward technique is to calculate the Stokes
vector produced along a given LOS in a magnetic model, and build an
image from a grid of sight lines. To do this, the magnetic field,
temperature, density, and velocity at every location along each LOS
are used. Given this information, we calculate the level populations
and the emitted polarization profiles for the Fe \noun{xiii} 1074.7
nm transition at each location using a publicly available Fortran
code ({\sf FORCOMP}) discussed by \inlinecite{Judge2001}. The forward model
has an IDL user interface and is publicly available for download and
use (\url{people.hao.ucar.edu/sgibson/FORWARD/}).

{\sf FORCOMP} first calculates the statistical-equilibrium equations based
on the location and the local plasma parameters from the model: height
above the solar surface [$h$], density [$\rho$], temperature [$T$], magnetic
field [$\vec{B}$], and velocity [$\vec{v}$]. Using standard atomic data,
the statistical equilibrium equations determine the relevant level
populations of the atomic system for the transition in question. The
code treats inelastic and superelastic collisional processes, but
neglects elastic collisions. This omission
affects the magnitude, but not the direction, of $L$ and leads to
a small uncertainty in $V$. The LOS field strength and the POS field
direction are not strongly affected by the elastic collisions. Once
the level populations are determined, {\sf FORCOMP} solves the radiative-transfer equations to calculate the polarization of the reemitted
radiation in the direction of the observer \cite{Judge2001}. The
signals are then integrated over wavelength into a single number for
each pixel (or LOS), and are assembled into an image.

The benefit of the forward technique is that we can easily calculate
the simulated polarization signals from a theoretical model of a magnetic
system and then compare these images with observations. It
allows us to test the theories against an observable that is directly
sensitive to the magnetic field in the corona. Additionally it allows
for comparison between the models themselves.

The forward model outputs Stokes $I$, $Q$, $U$, $V$ and combinations
thereof. We use mainly intensity [$I$],
relative linear polarization [$L/I$], azimuth [$\Psi$], and
relative circular polarization [$V/I$].

\section{MHD Models \label{sec:mhd}}

For our study of magnetic flux rope and sheared-arcade signatures in
the corona, we used three models, each having a distinct magnetic
morphology: The first model is a 3D analytic spheromak flux rope in exact equilibrium \cite{Gibson1998_giblow,gibson2000_giblow}.
The second is an azimuthally symmetric (2.5D) cylindrical flux rope taken from an MHD simulation created to study current-sheet formation during CME initiation \cite{fan200_yfan2d1}.
The last is a 2.5D sheared arcade taken from MHD simulations of 
CME initiation by the multipolar breakout mechanism
(\opencite{Antiochos1999}; \opencite{Karpen2012},  and references therein).
All of the models are in, or near, equilibrium and have been argued as models for prominence magnetic structure. The two flux-rope models contain a region of concave-up magnetic dips that can support prominence plasma against gravity, and they also capture many observed properties of  coronal cavities \cite{Hudson1999,gibson2000_giblow,Mackay2010,Reeves2012}. Sheared arcades in 2.5D typically contain only concave-down regions that can support time-dependent prominence condensations if the field is sufficiently flat \cite{Karpen2001}. In three dimensions, sheared arcades can develop regions of concave-up field lines like those in flux ropes, and they then can support the plasma statically against gravity (\opencite{Antiochos1994}; \opencite{Luna2012}, and references therein).

In the work presented here, we use a single snapshot from the two time-dependent MHD models. 
Because we are studying the steady-state pre-CME magnetic
structure, all velocities have been set to zero. The times used are those where
the field is near equilibrium, and thus have close to zero velocity everywhere.

Two of the models that we study are 2.5D. Azimuthal symmetry creates
structures that are elongated along the LOS. When structures are highly
3D, the magnetic information can become smeared along the LOS, making
magnetic signatures more difficult to identify. We use 2.5D models
because the observational signatures of the magnetic field we are
studying are clear coronal cavities. When a coronal filament channel and associated
neutral line are along the LOS -- nearly parallel with the solar equator -- a
cavity commonly becomes visible implying that cavities are elongated along the LOS \cite{gibson2010-cavities}. The 2.5D assumption is thus justified by cavity observations.

For the spheromak model, we used the density and temperature provided
by the analytic model. The parameters were chosen such that the density and temperature vary only slightly in the calculation domain \cite{Dove2011}. 
For the two MHD models, a range of plasma parameters was explored.
The goal of this work is to study the impact of the magnetic morphology
on the polarization signatures.  We looked at our models both with the original plasma
distributions from the MHD simulations and with simple spherically symmetric plasma profiles. These new plasma distributions are not strictly in equilibrium with the magnetic fields.
However, they serve the useful  purpose of providing a means of disentangling
those features in the polarization data that are due to the magnetic
morphology from those that are heavily influenced by the plasma parameters.
In addition, any effects of changing the plasma distributions on the magnetic field structures would be very small, since the prominence and cavity are in the low-$\beta$ regime and are nearly in force-free equilibrium in both MHD examples. 

Coronal Stokes vectors are calculated for each theoretical system
using the forward code described in Section \ref{sec:forward model}.
We compare these polarization signatures with each other to determine
their similarities and differences and to identify their distinguishing
features.

\subsection{Model Descriptions\label{sub:models}}

\begin{figure}  

   \centerline{\hspace*{0.015\textwidth}
     \includegraphics[width=0.5\textwidth,clip=]{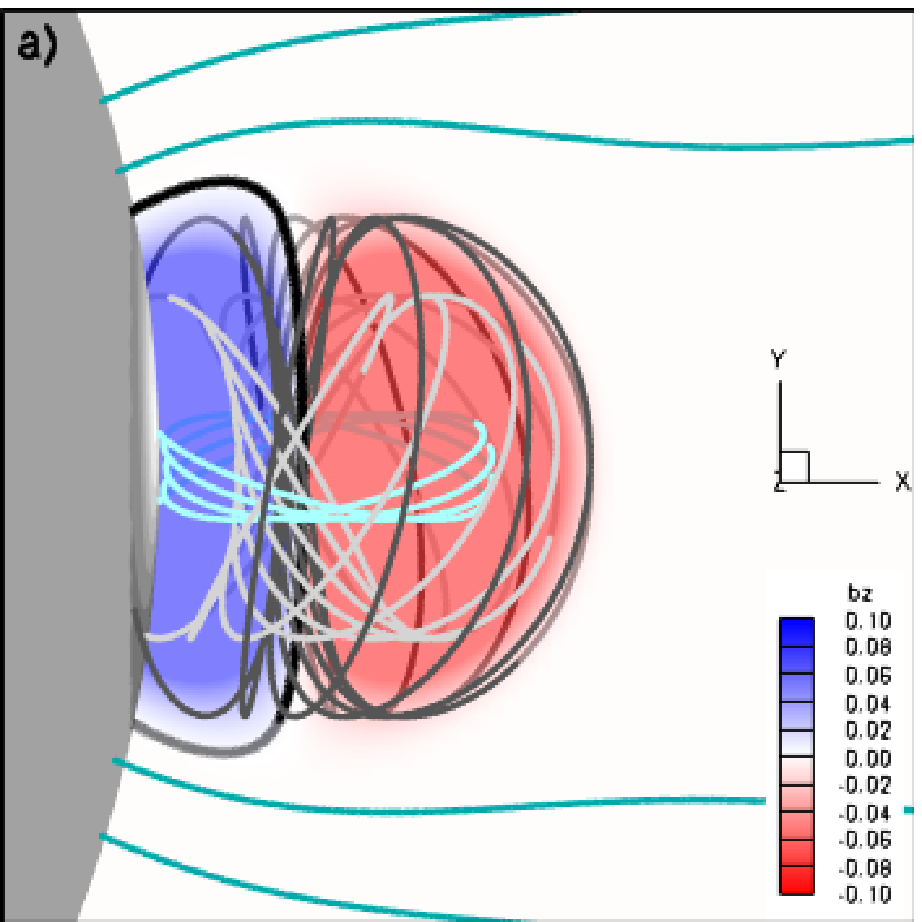}
               \includegraphics[width=0.5\textwidth,clip=]{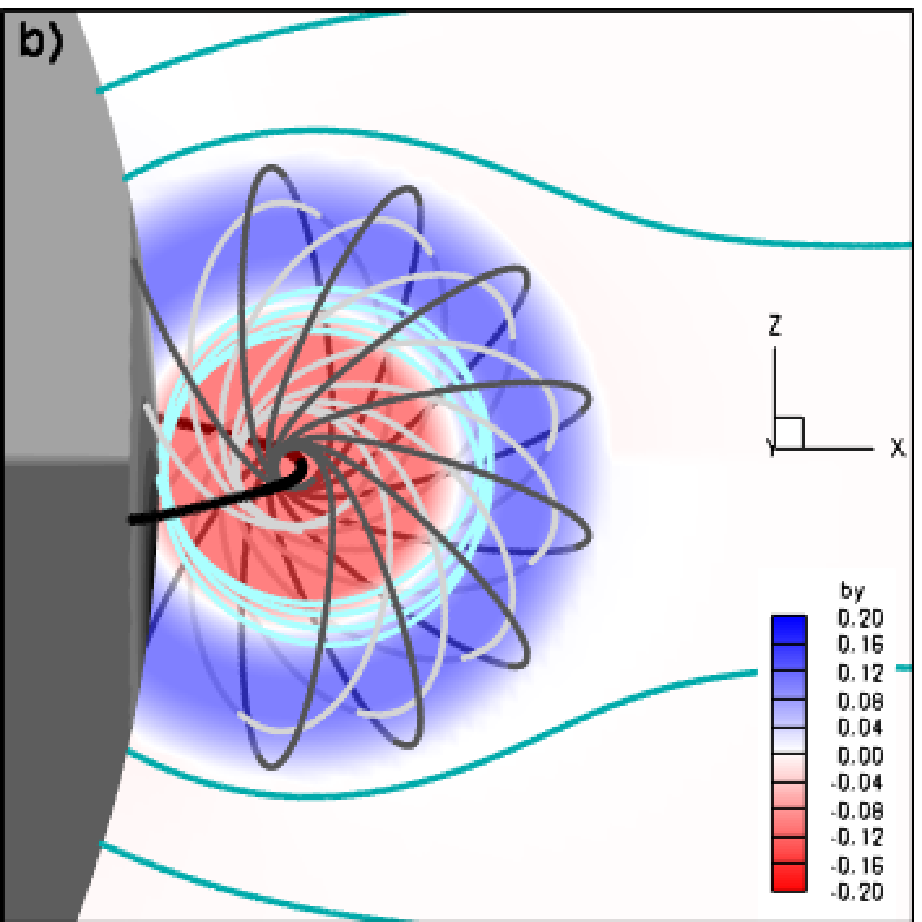}
              }
     
\caption{Field-line traces in the analytic spheromak flux-rope model 
seen from two different view points. Prominence material
would theoretically sit below the thick black line, which traces the
flux-rope axis, in the dips of the magnetic field. In the forward
calculations, we used the orientation in b), with the magnetic field along
the axis pointed toward the observer. An animated version of this figure is available in the electronic supplementary material.}
\label{fig:giblow}
\end{figure}

The first magnetic system we explored is that of an analytic spheromak
flux rope. A more detailed study of forward model results from this
particular flux rope can be found in \inlinecite{Dove2011}. The spheromak
model (Figure \ref{fig:giblow}) is an exact solution to the MHD equations
in full magnetostatic equilibrium \cite{Gibson1998_giblow}. The
magnetic field of the flux rope is a closed, twisted-flux system, attached
to the photosphere, which has been shown to reproduce observational
features of a three-part CME including the cavity and the bright prominence
\cite{gibson2000_giblow}. The external field has a split bipolar configuration with a hydrostatic density background. We used
an orientation such that the flux-rope axis, and hence the prominence
material, is oriented along the LOS, and the axial magnetic field is
directed toward the observer (Figure \ref{fig:giblow}(b)). As stated by \inlinecite{Dove2011}, we chose a parameter set such that the density
is close to spherically symmetric. The background-density profile
was taken from \inlinecite{Schmit2011}. The density decreases from around
$5\times10^{8}$~cm$^{-3}$ at photosphere to about $3\times10^{7}$~cm$^{-3}$ near the top of the spheromak at 1.3~R$_{\odot}$. The
temperature is between $7\times10^{5}$ and $1\times10^{6}$~K. The
magnetic-field strength is strongest at the axis where it is around
$1$~G, and the external field strength near the flux rope is of order $0.1$~G. Thus, the plasma $\beta$ is high outside the spheromak, above $100$,
and between $1$ and $10$ in most of the flux rope.

\begin{figure}  

   \centerline{\hspace*{0.015\textwidth}
     \includegraphics[width=0.7\textwidth,clip=]{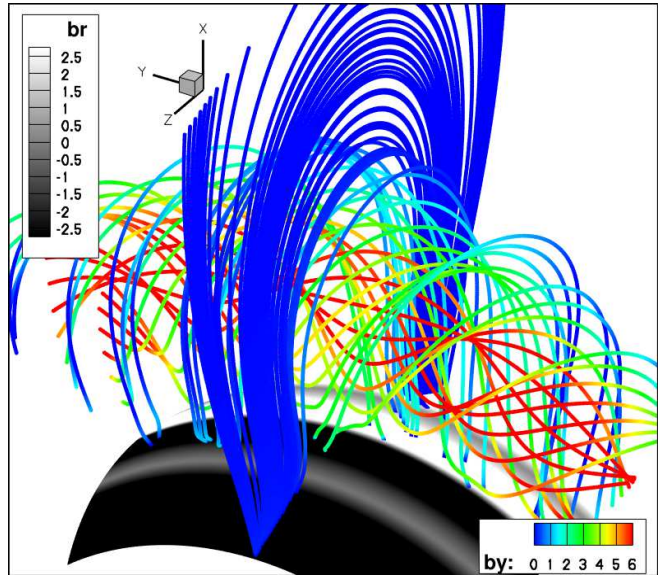}
              }
\caption{Field lines traced from the 2.5D axisymmetric cylindrical flux rope. Colors on the field lines
represent the strength of the $y$, or axial, component of the field. An animated version of this figure is available in the electronic supplementary material.}
\label{fig:yfan} 
\end{figure}

The second model is a 2.5D axisymmetric cylindrical flux rope (Figure
\ref{fig:yfan}) and is described in \inlinecite{fan200_yfan2d1}. This
numerical model comprises a $2\times10^{6}$ K isothermal atmosphere
occupied by a potential arcade field under which a twisted toroidal
flux tube is kinematically emerged. With continual emergence, the
system is quasi-static until time $t$ = 114~R$_{\odot}/v_{\mathrm A0}$ where
$v_{\mathrm A0}$ is a characteristic Alfv\'en speed, after which the
flux rope erupts. If the emergence is stopped before that, the system
remains stable. We analyzed a time step at $t = 114$
where the emergence was stopped at $t=112$. The model extends from 1~R$_{\odot}$ to 14.4~R$_{\odot}$ radially and from $\pi/3$ to $2\pi/3$
in latitude. For the forward analysis, the system is oriented such that flux rope 
is in the equatorial plane and the axial field points 
away from the observer. The magnetic-field strength at the axis is around $10$ G.
The electron density in the flux rope is on the order of
$10^{6-7}$~cm$^{-3}$. The plasma $\beta$ is below $0.1$ except near the 
footprint of the arcade immediately surrounding the flux rope.

\begin{figure}  

   \centerline{\hspace*{0.015\textwidth}
     \includegraphics[width=0.7\textwidth,clip=]{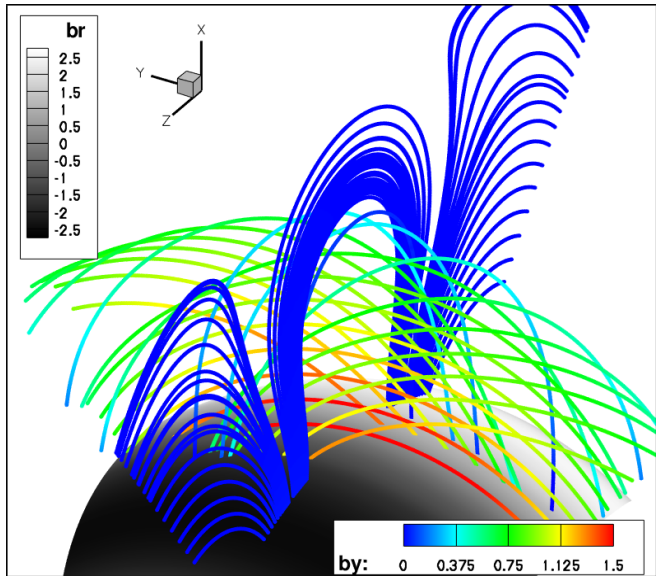}
              }
\caption{Field lines traced from the 2.5D sheared arcade model. Colors
on the field lines represent the strength of the $y-$, or axial, component
of the field. An animated version of this figure is available in the electronic supplementary material.}
\label{fig:devore-fieldlines}
\end{figure}

The third model (Figure \ref{fig:devore-fieldlines}) is a 2.5D axisymmetric numerical 
datacube of a breakout
quadrupolar system \cite{Karpen2012}. The computational domain
extends from $1$ to $125$~R$_{\odot}$ radially and over $\pi$ radians
in latitude. The system is energized by shearing the innermost polarities 
near the neutral line. The shearing results in a field that is pointed away 
from the observer in our orientation. 
The field that connects these two polarities shears and expands both
radially and laterally. Because a simple adiabatic energy equation and
a closed lower boundary condition are used, the plasma entrained in the 
sheared field rarefies and adiabatically cools relative to the background.
This configuration does not form a flux rope with a central axis until flare reconnection
sets in following the eruption.  The time shown in Figure \ref{fig:devore-fieldlines} 
is $t = 60\,000$~seconds; stopping the shearing motions at this time results in a 
stable equilibrium state, while continuing the motions leads inexorably 
to an eruption. In and around the sheared arcade, the magnetic-field 
strength is around $1$~G. In this same area, the plasma densities 
are $10^{6-7}$~cm$^{-3}$. The temperature reaches a local minimum around 
$3\times10^{5}$~K within the sheared field, while it is about $10^{6}$ in the 
surrounding unsheared field. The plasma $\beta$ is below $0.01$ in the sheared 
region, and is of order unity in the unsheared region (except of course, in the 
immediate vicinity of the null point, where the $\beta$ becomes very large).

\section{Results\label{sec:Results}}

Interpretation of a polarization image is not necessarily straightforward
(see \opencite{Judge2007}; \opencite{Rachmeler2012} for detailed descriptions
of the signal interpretation). For $L/I$ images ({\it i.e.} Figure \ref{fig:giblow_results}(b)),
the magnitude of the signal is always positive, and the images are
usually plotted on a logarithmic scale. Bright areas indicate
magnetic field that is primarily in the POS. Dark areas generally
indicate magnetic field that is along the LOS. Sharp, elongated dark
structures are usually indicative of Van Vleck inversions, marking
where the magnetic field is at an angle of $\approx54^{\circ}$ from
radial. The $L/I$ images may have magnetic-field direction overlaid
as arrows or lines. In the images presented in this article, the red
arrows indicate the true POS direction of the magnetic field in a
thin POS slice that bisects the Sun. The blue lines indicate the azimuth
direction of the linear polarization, which is subject to the $90^{\circ}$
Van Vleck ambiguity (Section \ref{sub:Coronal-Stokes-vector}). Notice
that they are generally parallel when the magnetic field is close
to radial, and perpendicular when it is not. $V/I$ images are
plotted on a linear scale with blue as negative and red as positive;
white is zero. In our coordinate system, positive Stokes $V$ (blue)
indicates magnetic field toward the observer.

\subsection{Spheromak Flux Rope}

\begin{figure}  

   \centerline{\hspace*{0.015\textwidth}
     \includegraphics[width=0.4\textwidth,clip=]{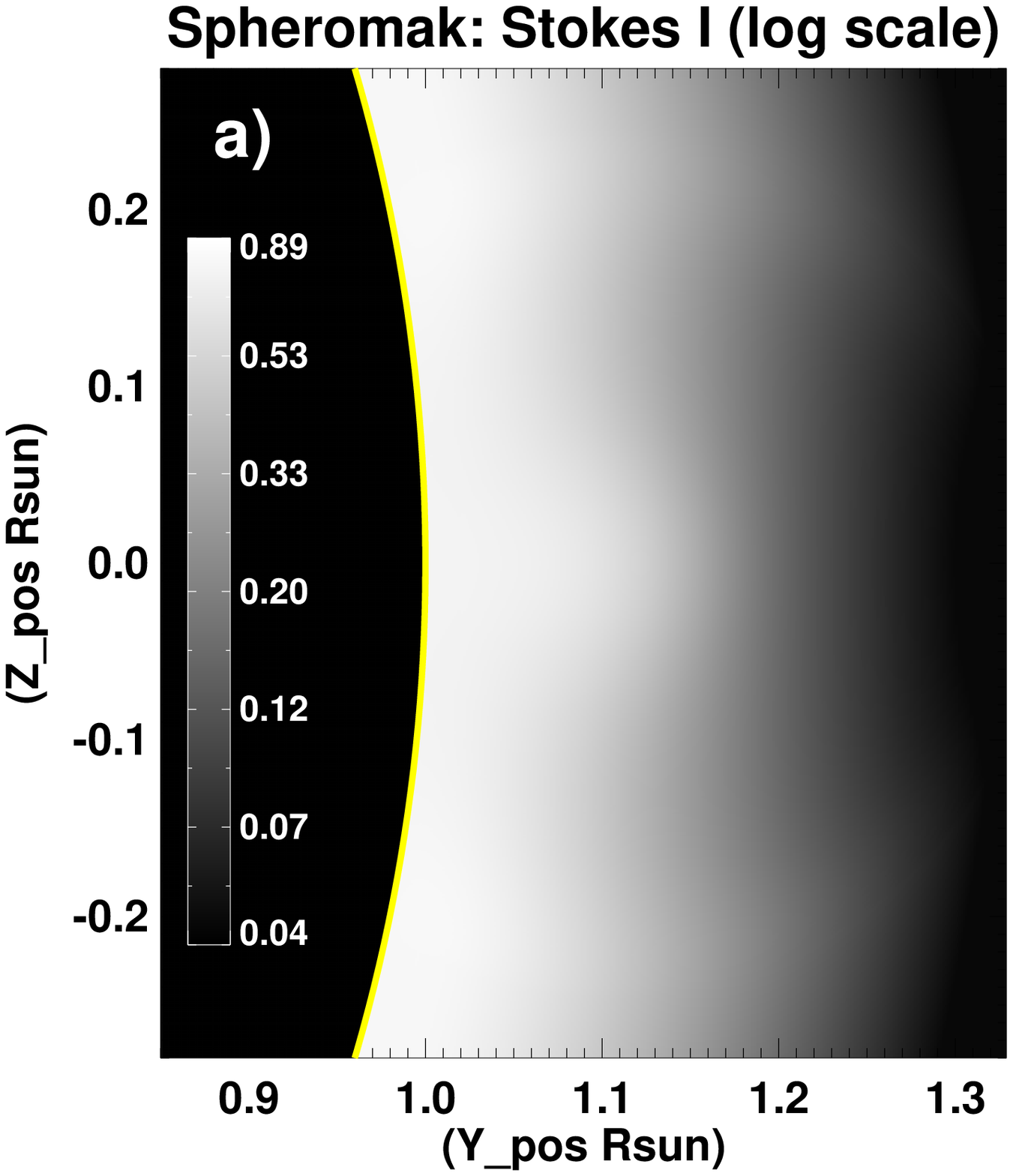}
               \includegraphics[width=0.4\textwidth,clip=]{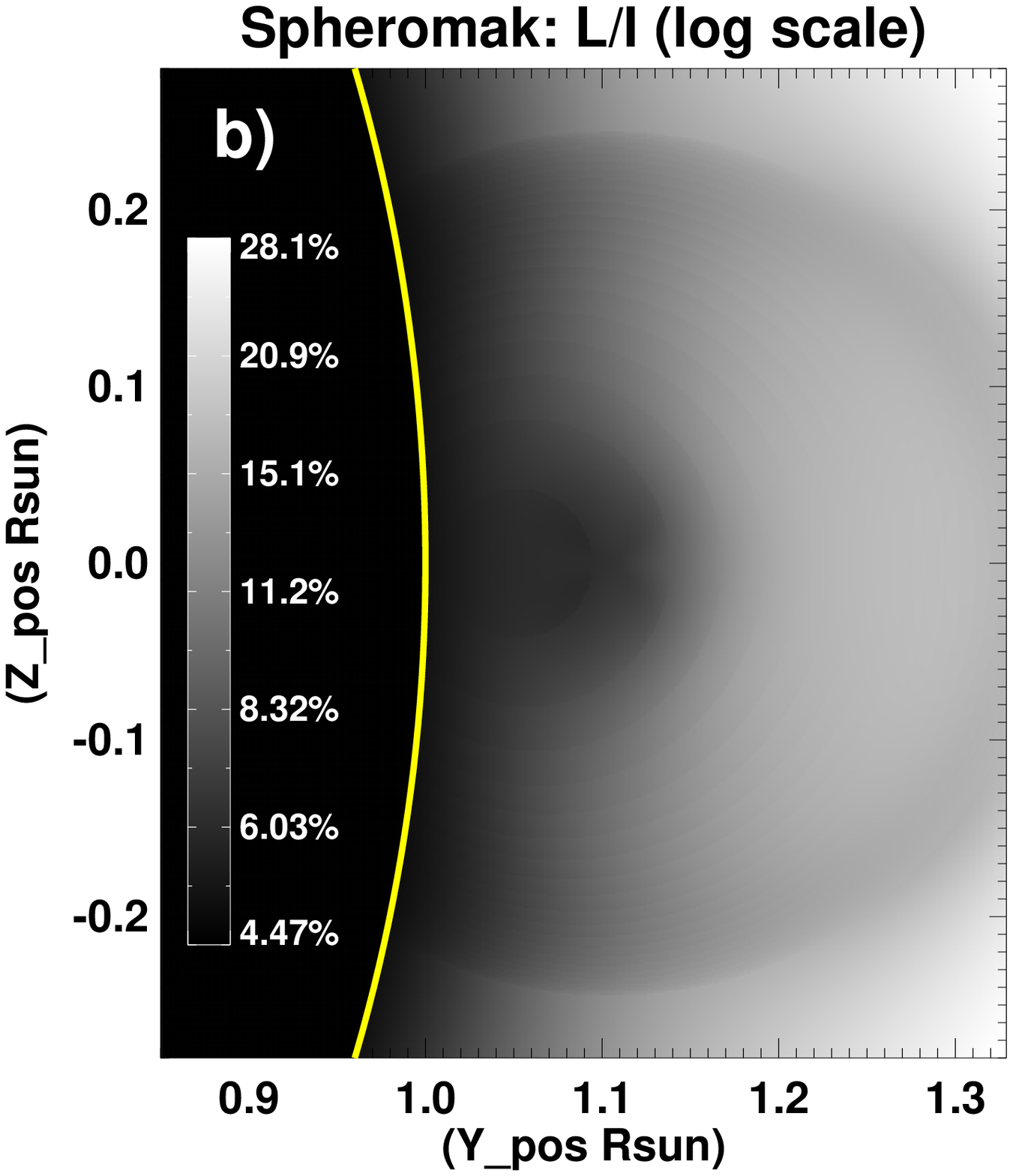}
              }
              \vspace{0.01\textwidth}
   \centerline{\hspace*{0.015\textwidth}
     \includegraphics[width=0.4\textwidth,clip=]{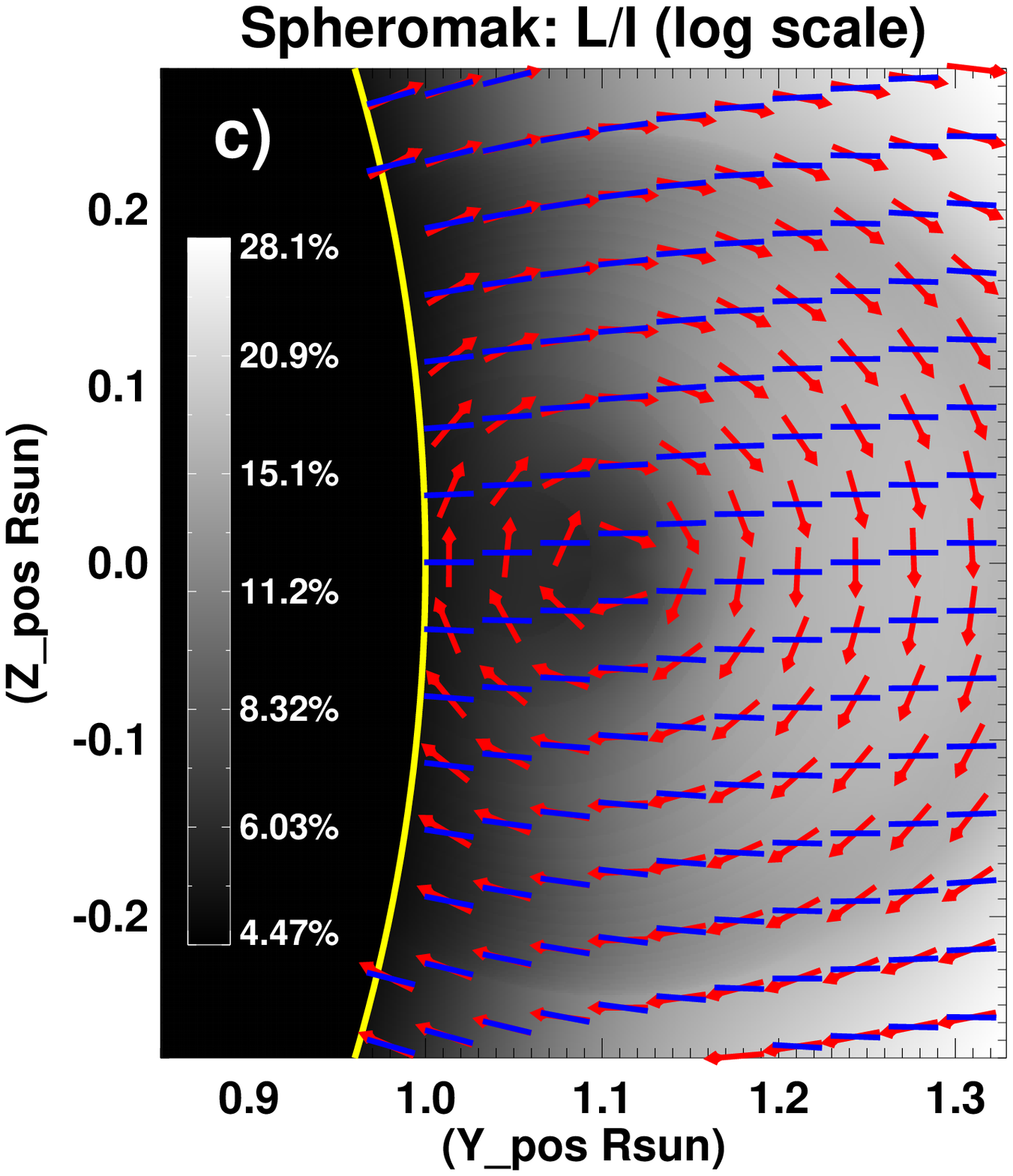}
               \includegraphics[width=0.4\textwidth,clip=]{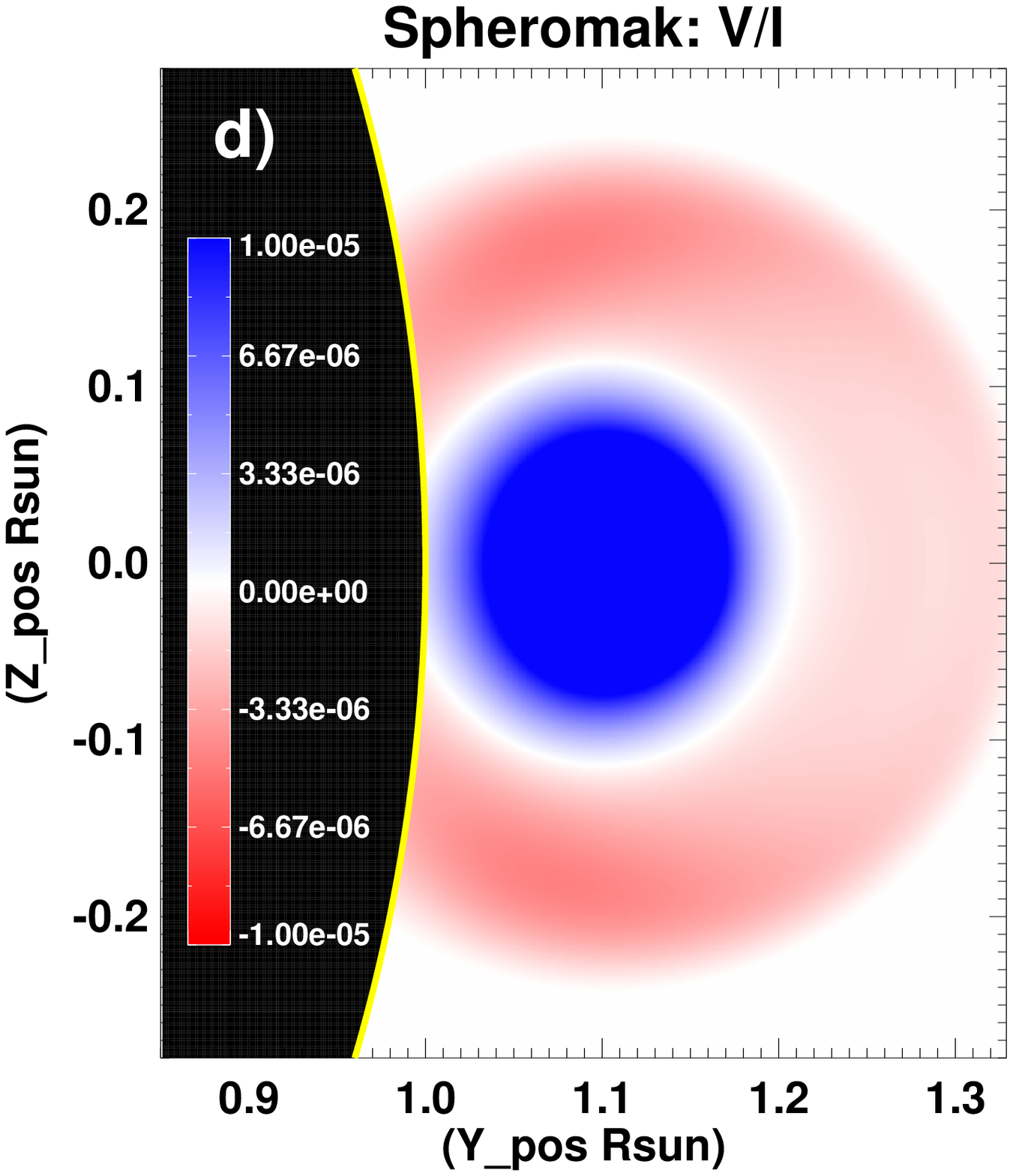}
              }
     
\caption{Forward-model results of the spheromak flux rope. a) Stokes intensity,
b) relative linear polarization, c) relative linear polarization with
magnetic-field direction plotted as red arrows and integrated polarization
azimuth direction plotted as blue lines, d) relative circular polarization.}
\label{fig:giblow_results}
\end{figure}

We use the parameter set from \inlinecite{Dove2011} to demonstrate the
main features for the spheromak flux rope. In the LOS integrated images,
the following features are identifiable and are robust signatures
of this magnetic morphology. Figure \ref{fig:giblow_results} shows
the forward model results, as presented also in \inlinecite{Dove2011}.
We summarize the conclusions from that analysis as follows: 
\begin{enumerate}
	\item \emph{Dark $L/I$ central core.} A dark core is clearly visible at
the location of the flux-rope axis in the $L/I$ image 
(Figure \ref{fig:giblow_results}(b); $Z = 0$, $Y = 1.1$). This is due
to the LOS field associated with the axis. The axis of this type of
flux rope is particularly clear because it is straight along the LOS,
and not curved like the axis of the cylindrical flux rope. The dark
central core is visible if the axis is oriented within about $30^{\circ}$
of the LOS. 
	\item \emph{Dark $L/I$ outer ring.} A ring of darker $L/I$ is visible
at the edge of the spheromak bubble (Figure \ref{fig:giblow_results}(b)), 
and it is also associated with
LOS field. This ring is much fainter than the axis field as the $\mathrm{\mathbf{B}}$
on the outer edge of the bubble is only aligned with the LOS in a
relatively narrow volume of space. 
	\item \emph{Bright $L/I$ ring.} Between i) and ii) is a bright
ring in $L/I$ (Figure \ref{fig:giblow_results}(b)), which is due to the POS field 
in the flux rope. 
	\item \emph{Radial azimuth.}The linear polarization direction shows no
clear Van Vleck inversion locations (blue lines in Figure \ref{fig:giblow_results}(c)). 
Although there are Van Vleck
inversions within the spheromak, the rotation of the field along the
LOS smears these out such that they are not visible in the integration. 
	\item \emph{Bi-directional circular polarization ($V/I$). }The circular
polarization comprises a clear circular positive signal around the axis surrounded by a weaker rung of negative signal (Figure \ref{fig:giblow_results}(d)).
The presence of both positive and negative Stokes $V$ is not found
in either of the other models studied here.
\end{enumerate}

\subsection{Cylindrical Flux Rope}

\begin{figure}  

   \centerline{\hspace*{0.015\textwidth}
     \includegraphics[width=0.4\textwidth,clip=]{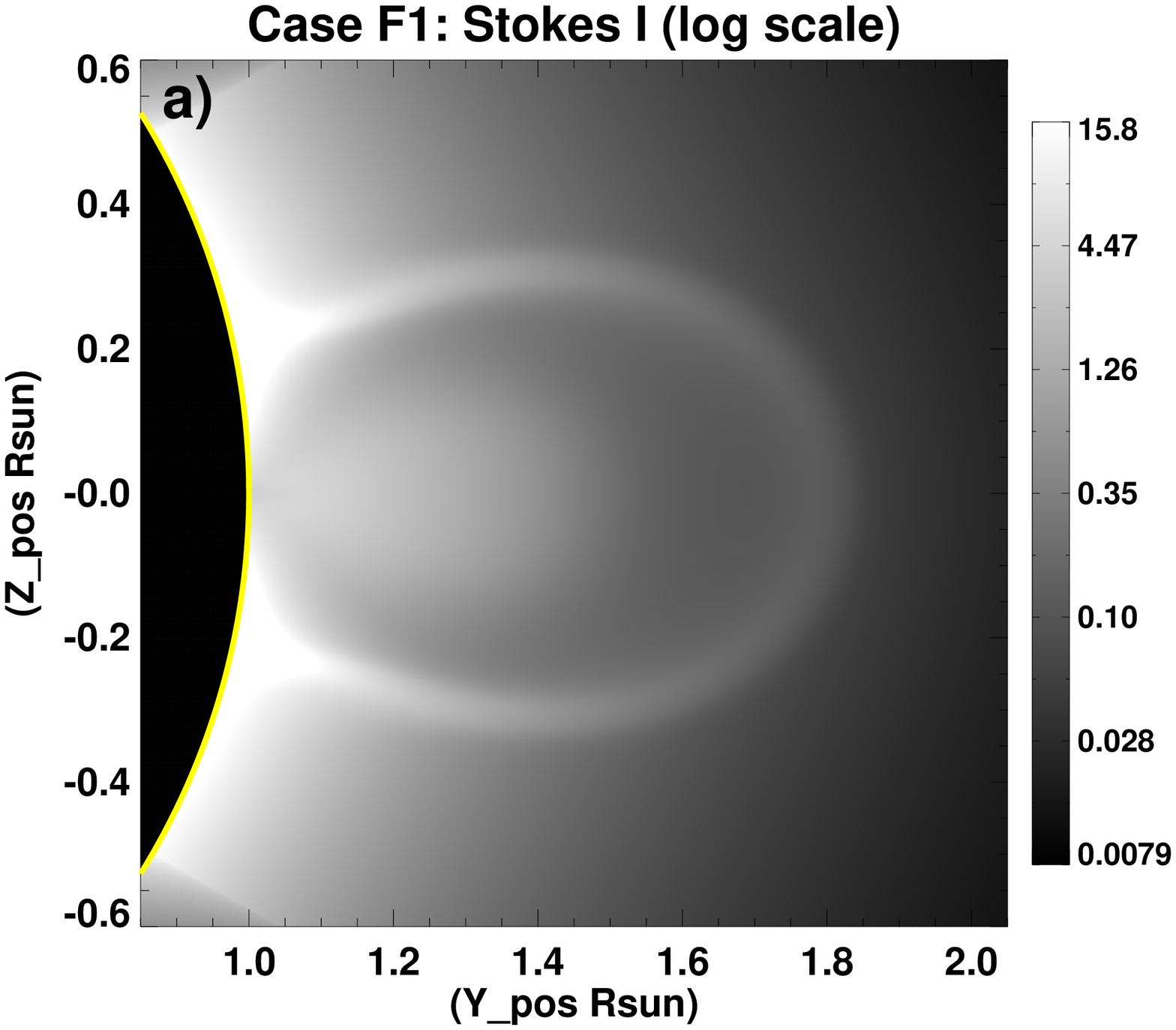}
               \includegraphics[width=0.4\textwidth,clip=]{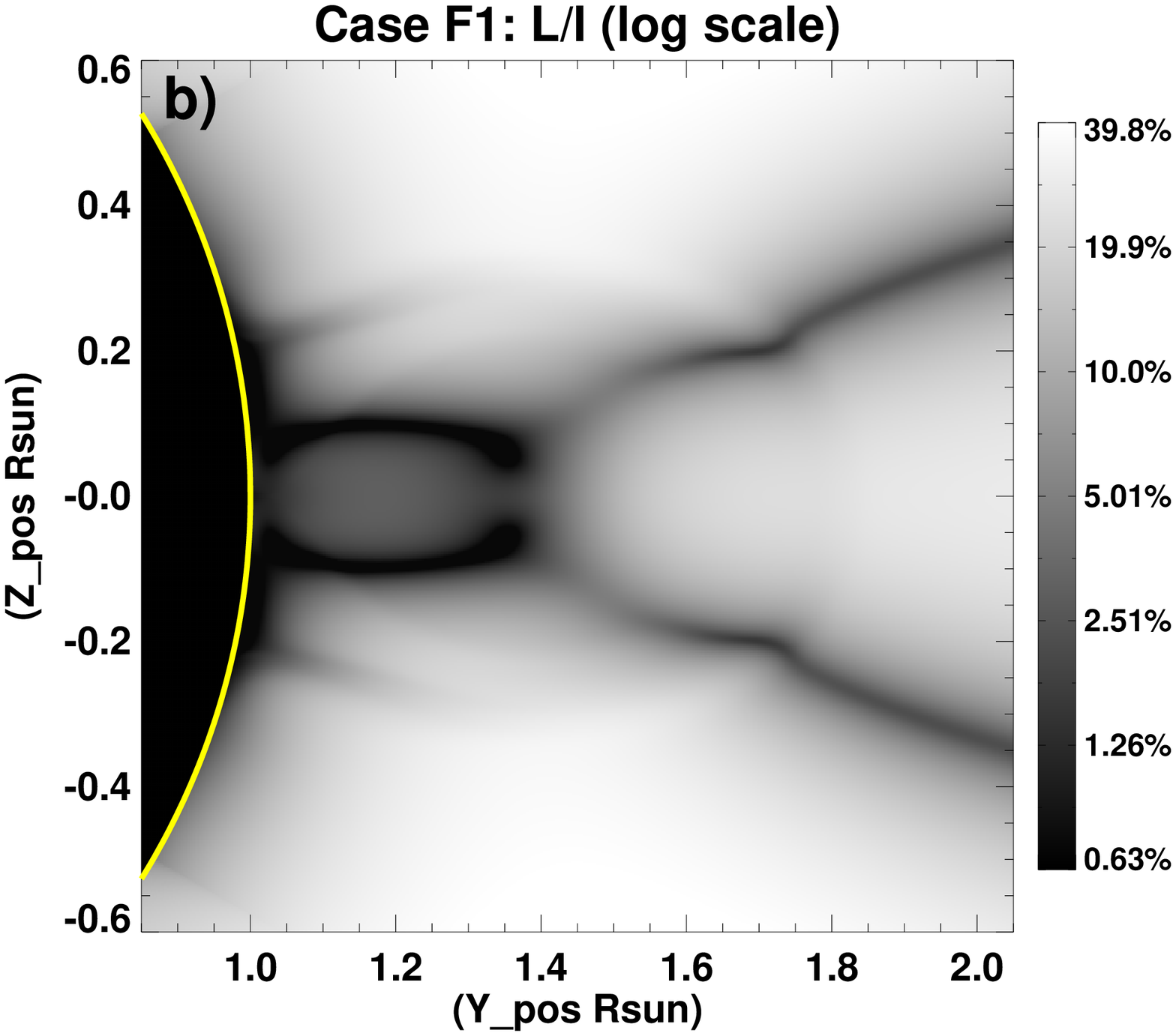}
              }
              \vspace{0.01\textwidth}
   \centerline{\hspace*{0.015\textwidth}
     \includegraphics[width=0.4\textwidth,clip=]{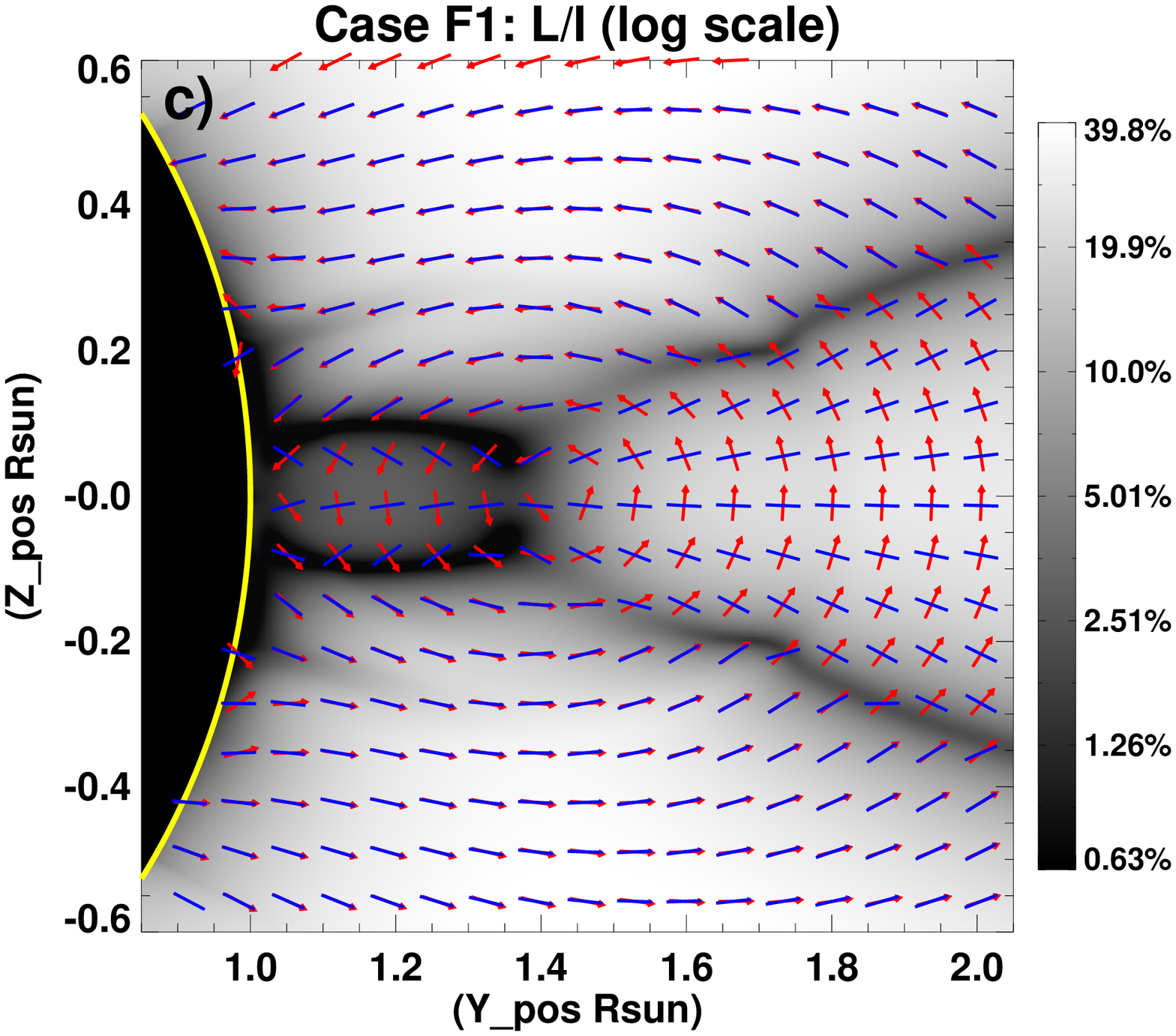}
               \includegraphics[width=0.4\textwidth,clip=]{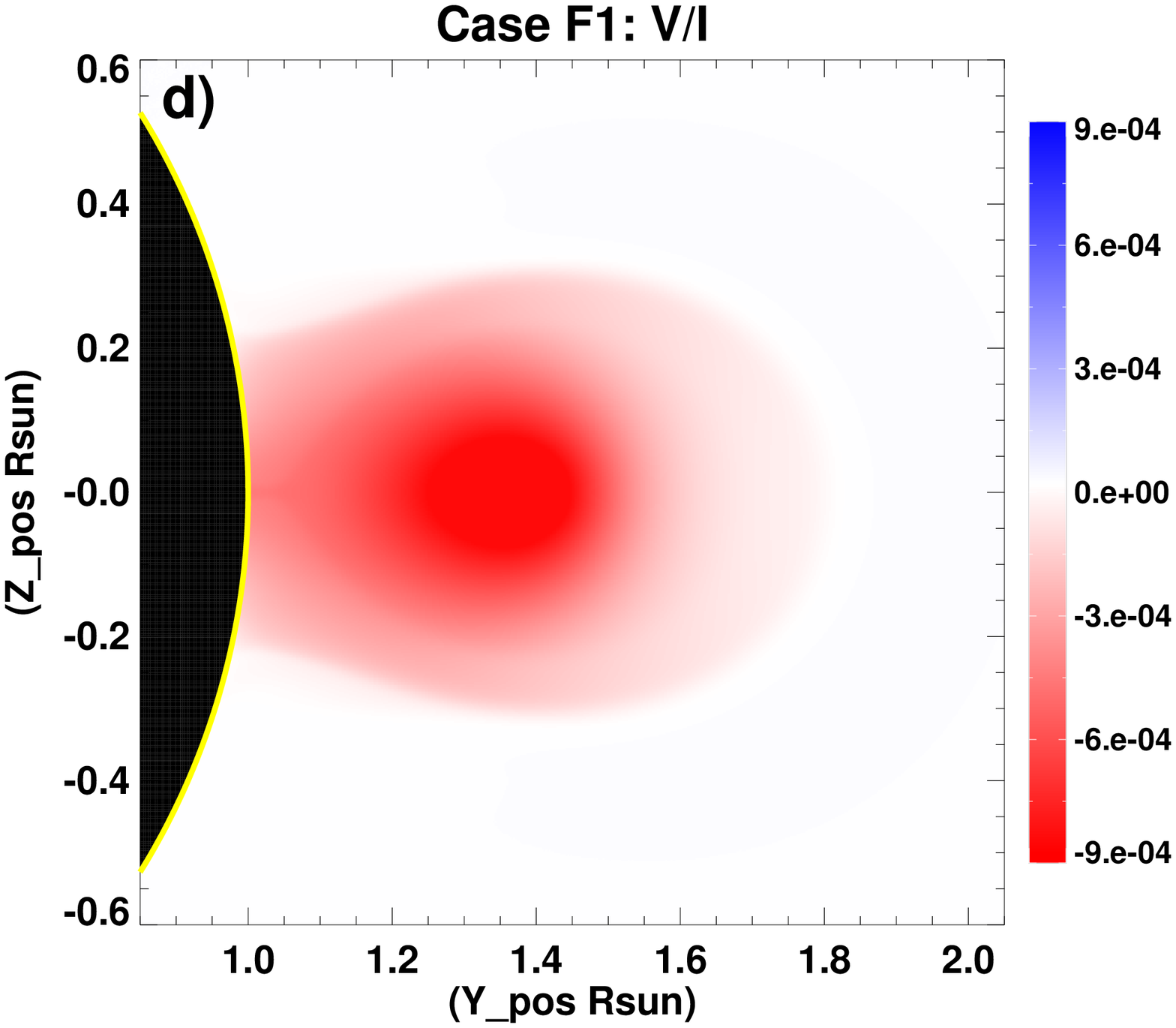}
              }
     
\caption{Forward-model results from the flux rope
model case F1 with the density of the original MHD data cube and a
temperature of $2\times10^{6}$~K. a) Stokes intensity, b) relative
linear polarization, c) relative linear polarization with magnetic-field direction plotted as red arrows and integrated polarization
azimuth direction plotted as blue lines, d) relative circular polarization.}
\label{fig:yfan_results}
\end{figure}

Figure \ref{fig:yfan_results} shows the forward-model results for
the cylindrical flux-rope configuration. To test the robustness of
the magnetic signatures in the polarization signals, we ran the forward
calculations on three cases. Case F1 uses the original density
and isothermal temperature of $T=2\times10^{6}$ K from the MHD model
(Figure \ref{fig:yfan_results}). Case F2 uses an isothermal
hydrostatic density fall-off with a scale height of 
(2R$_{\odot}k_{b}T$)/($G$M$_{\odot}m_{p}$)$~\thickapprox~$0.13R$_{\odot}$ (R$_{\odot}$ and M$_{\odot}$ are the solar radius and mass, $k_{b}$
is Boltzman's constant, $G$ is the universal gravity constant, $m_{p}$
is the mass of a proton, and $T=1.5\times10^{6}$~K is the temperature)
and a density at the coronal base of $5.8\times10^{8}$~electrons~cm$^{-3}$.
Case F3 uses a hydrostatic power-law density and temperature
function derived from fits to coronal-streamer densities \cite{Gibson1999}.  
The magnetic field in this model is in nearly force-free equilibrium. The 
imposed plasma is also low-$\beta$. If re-relaxed to true equilibrium, the 
scale-height of the plasma along field lines would be altered, but the field 
topology would remain nearly unchanged. 

\begin{figure}  

   \centerline{\hspace*{0.015\textwidth}
     \includegraphics[width=0.3\textwidth,clip=]{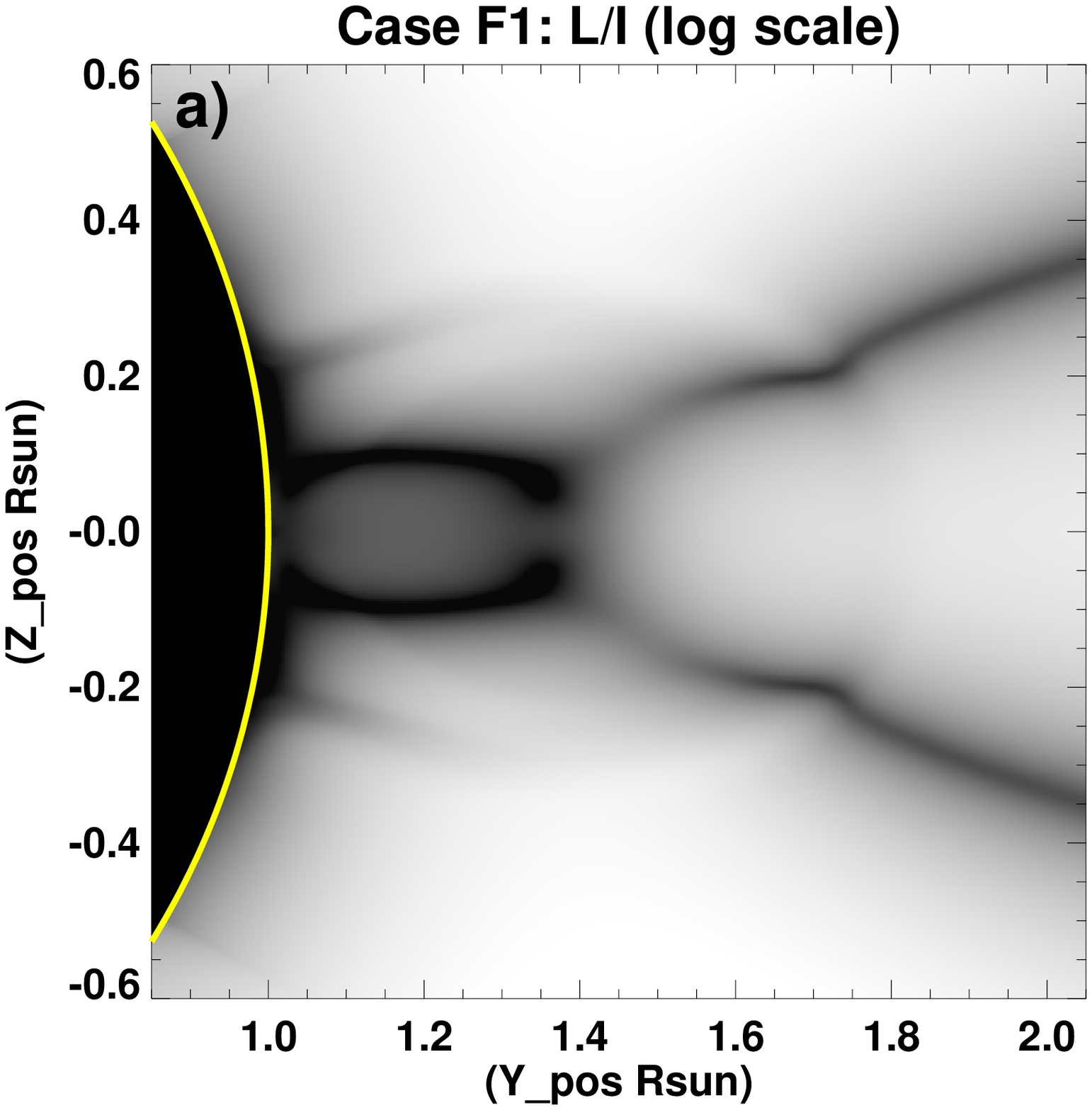}
               \includegraphics[width=0.3\textwidth,clip=]{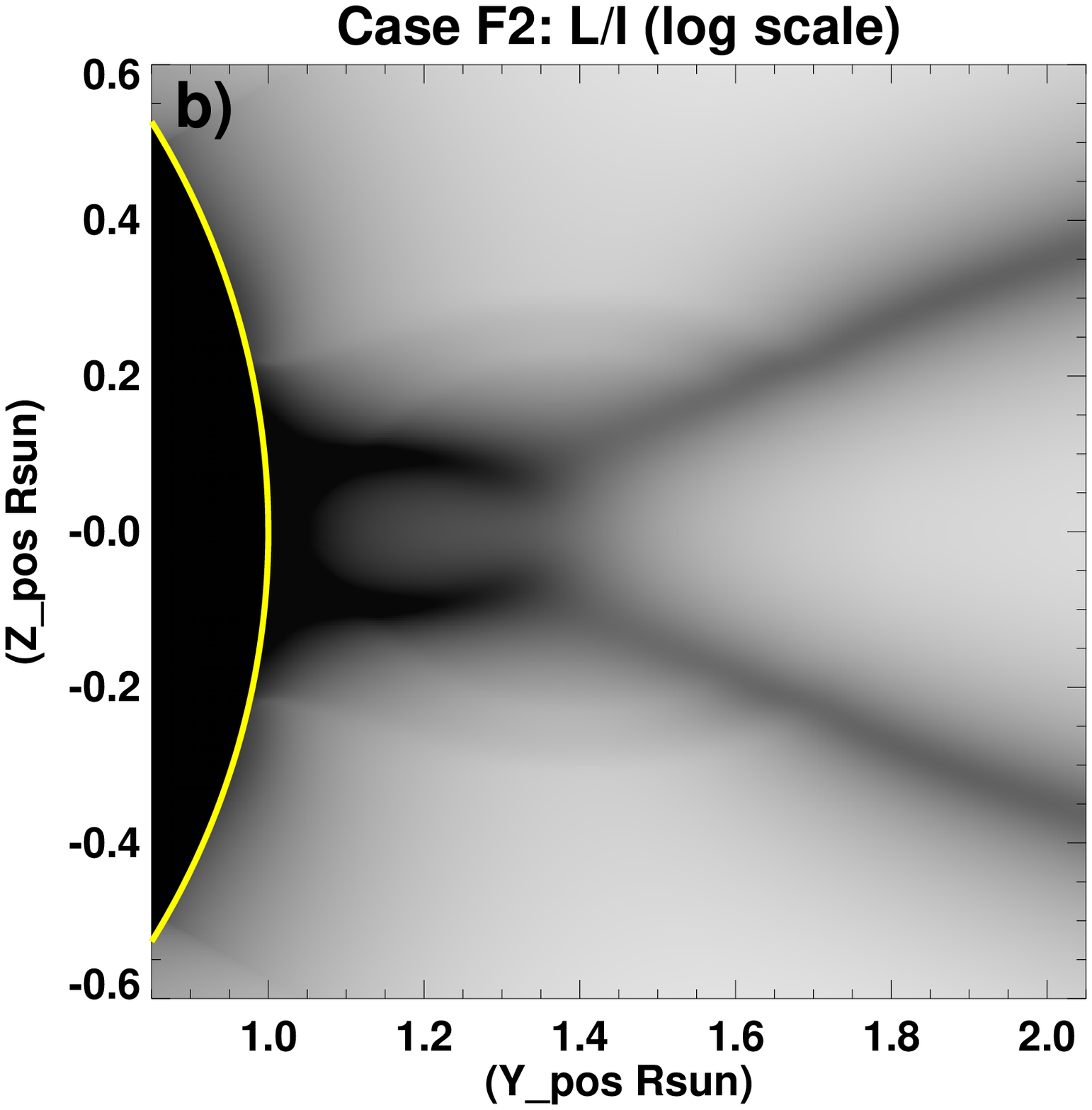}
               \includegraphics[width=0.35\textwidth,clip=]{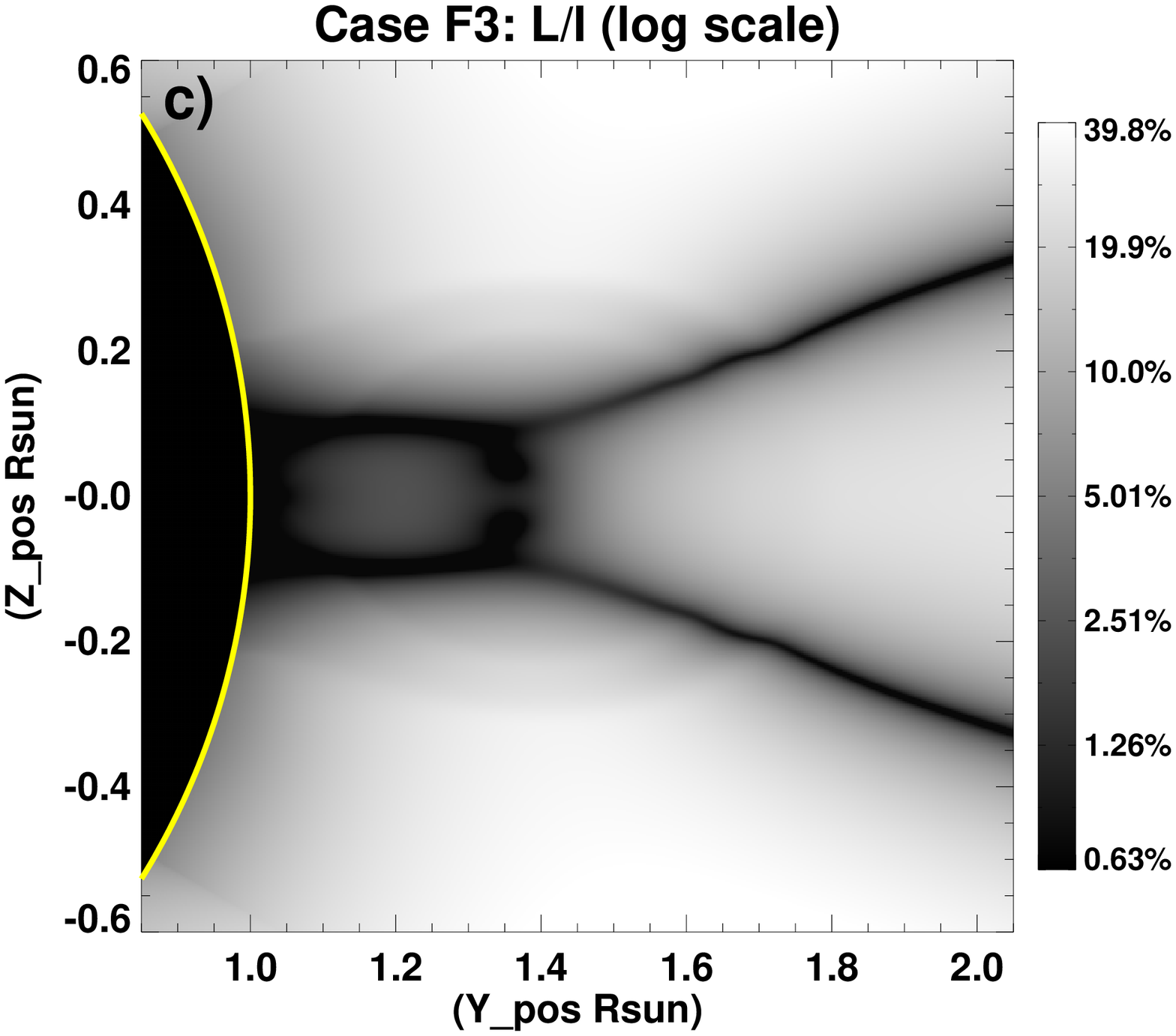}               
              } 
     
\caption{Comparison of the relative linear polarization for the three cases
of the cylindrical flux rope. All three images use the same scale.}
\label{fig:yfan-comparison}
\end{figure}

Stokes $I$ changed noticeably when a spherically symmetric density
was used. Almost none of the structure seen in F1 (Figure~\ref{fig:yfan_results}(a))
is present in F2 or F3 (not shown). This is not surprising as the
intensity of emitted radiation is strongly dependent on the local
plasma density, so a spherically symmetric density results in a virtually
spherically symmetric Stokes $I$. The relative linear polarization
signals had little variation between the three cases (Figure
\ref{fig:yfan-comparison}); these differences are discussed at the end of this Section.
We will first analyze the signatures of F1. 

\begin{enumerate}
	\item \emph{Dark $\vee$-shaped Van Vleck inversions in the arcade.} These are the Van
Vleck inversions in the external arcade field (Figure
\ref{fig:yfan_results}(b); $Y = 1.4$~to~$2$). The field outside of
the $\vee$ is less than $54^{\circ}$ from radial, and the field
inside the $\vee$ is greater than that. This is clearly visible (Figure
\ref{fig:yfan_results}(c)) in the shift of the linear-polarization
direction (blue lines) from parallel to perpendicular to the POS magnetic-field direction (red arrows).
These two Van Vleck inversions extend downward into the flux rope. 
	\item \emph{Darker central region in L/I.} In general, the region near the flux-rope axis ($Z = 0$, $Y = 1.35$) is slightly darker (Figure \ref{fig:yfan_results}(b)). This
is because there is more LOS field in this region, so there is less
linear-polarization signal. However, because of the LOS integration
and the curvature of the central flux-rope axis, the location of the axis is unclear. See also iii) below.
	\item \emph{Dark beads in L/I near the axis.} These are visible in F1 and F3, but
not in F2 (Figure \ref{fig:yfan-comparison}). These locations are dark because
they are signatures of the LOS field in the flux rope axis.
This flux rope is axisymmetric, so it curves around the limb of the Sun.
The curvature results in a perspective effect such that the location
of LOS field is only co-spatial with the location of the flux rope
axis in the central POS slice. The true location of the flux-rope axis
is not readily apparent in the integrated data because the volume of space where they
are co-located is a small. The location of the
LOS field follows a $\supset$-shaped arc whose legs coincide with
iv). 
	\item \emph{Two dark horizontal lines in L/I intersecting the photosphere.} These
are a combination of a Van Vleck inversion in the lower part of the
flux rope and LOS field from the legs of the $\supset$-shape described
in iii) and are thus the darkest features in the image. 
	\item \emph{Slightly spoked azimuth direction}. The azimuth directions are
mostly radial, but there is a slight spoke-like signature around the
flux-rope axis (Figure~\ref{fig:yfan_results}(c)). Since the Van Vleck
inversion locations are obvious (in this model), the $90^{\circ}$
ambiguity can be removed, and the flux rope nature of the field becomes
evident. Even without removing the ambiguity, the slight spoke may
be a feature that can help to identify this type of magnetic
morphology. Note that the linear-polarization azimuth direction (blue
lines) everywhere is close to radial.
	\item \emph{Bulb of circular polarization}. The circular polarization is
all the same sign (Figure \ref{fig:yfan_results}(d)). The strongest
signal comes from above the limb, and surrounds the location of the
flux rope axis.
\end{enumerate}
Although almost all of the listed features are present in each case,
some are more pronounced in certain cases. For instance, the dark
beads in iii) are distinctly visible in F3, and not at all in F2.
The density differences in each case change the weighting of
the signal along the LOS. Thus, certain features are more or less clear
depending on whether the signal is concentrated at the central POS, or
spread out along the LOS. Case F2 has the most gradual density
drop with height and thus the dark beads from iii) are overcome
by brighter signal in the foreground and background.

\subsection{Sheared Arcade}

Much like the cylindrical flux-rope model, we ran the forward calculations
on the sheared arcade model with several different density and temperature
profiles. The cases presented here are as follows: S0 -- the
density and temperature provided by the MHD model; S2 -- the
density provided by the MHD model and isothermal temperature of $1.5\times10^{6}$~K; S3 -- hydrostatic streamer density and temperature fit
from \inlinecite{Gibson1999}. We do
not relax the configuration to equilibrium with the imposed plasma
parameters. In all cases, the plasma is low-$\beta$ except in the region 
of the null-line. We will not discuss results from S0. The temperature of the original
data falls below the minimum temperature threshold for the forward
calculations (around $5\times10^{5}$ K). Most of the sheared-field
plasma is at or below this temperature, due to the assumption of 
adiabatic energy transport, so any calculations
on the original data only produce signal from the unsheared field,
which is not useful for this work. 

\begin{figure}  

   \centerline{\hspace*{0.015\textwidth}
     \includegraphics[width=0.4\textwidth,clip=]{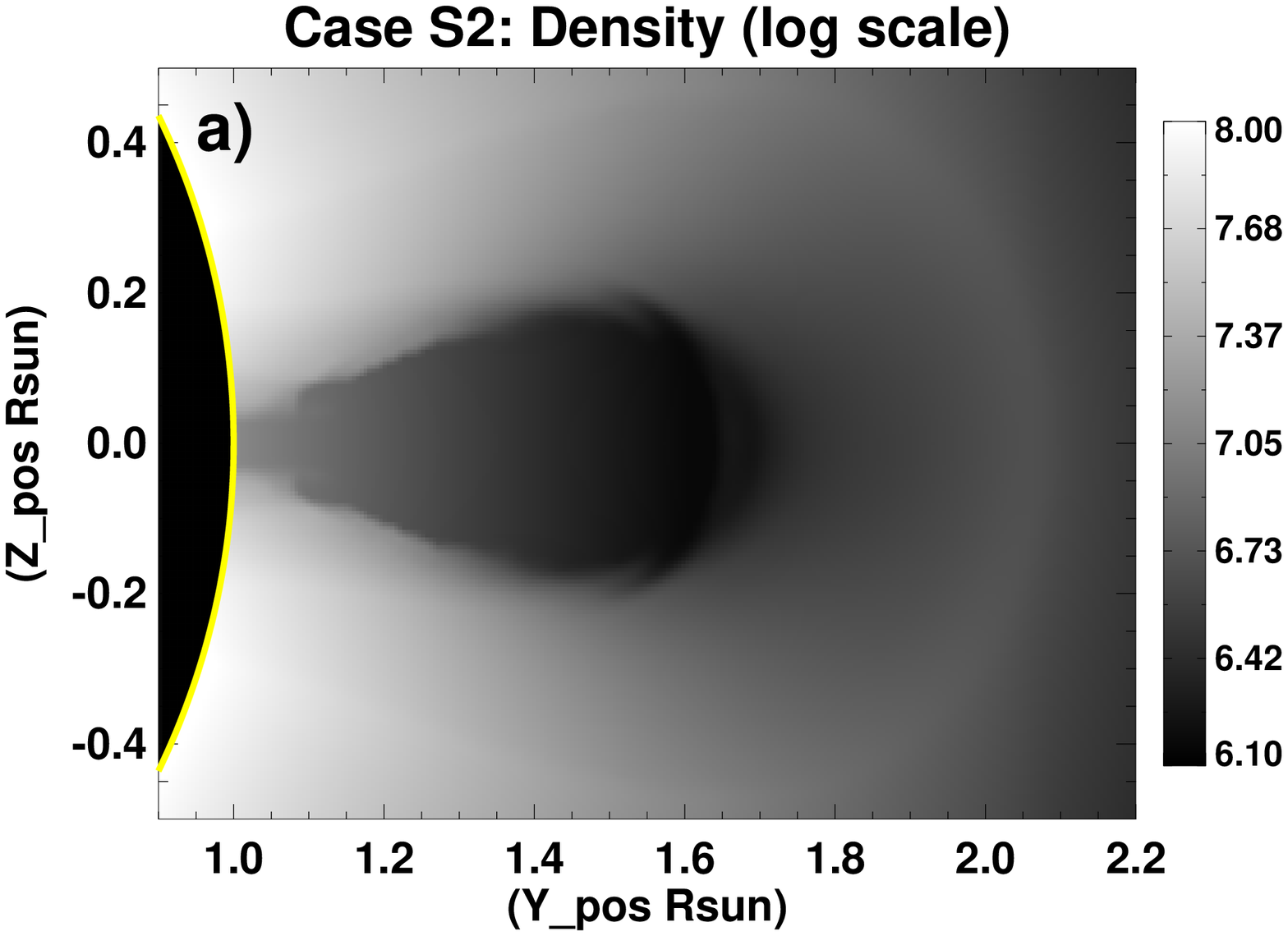}
               \includegraphics[width=0.4\textwidth,clip=]{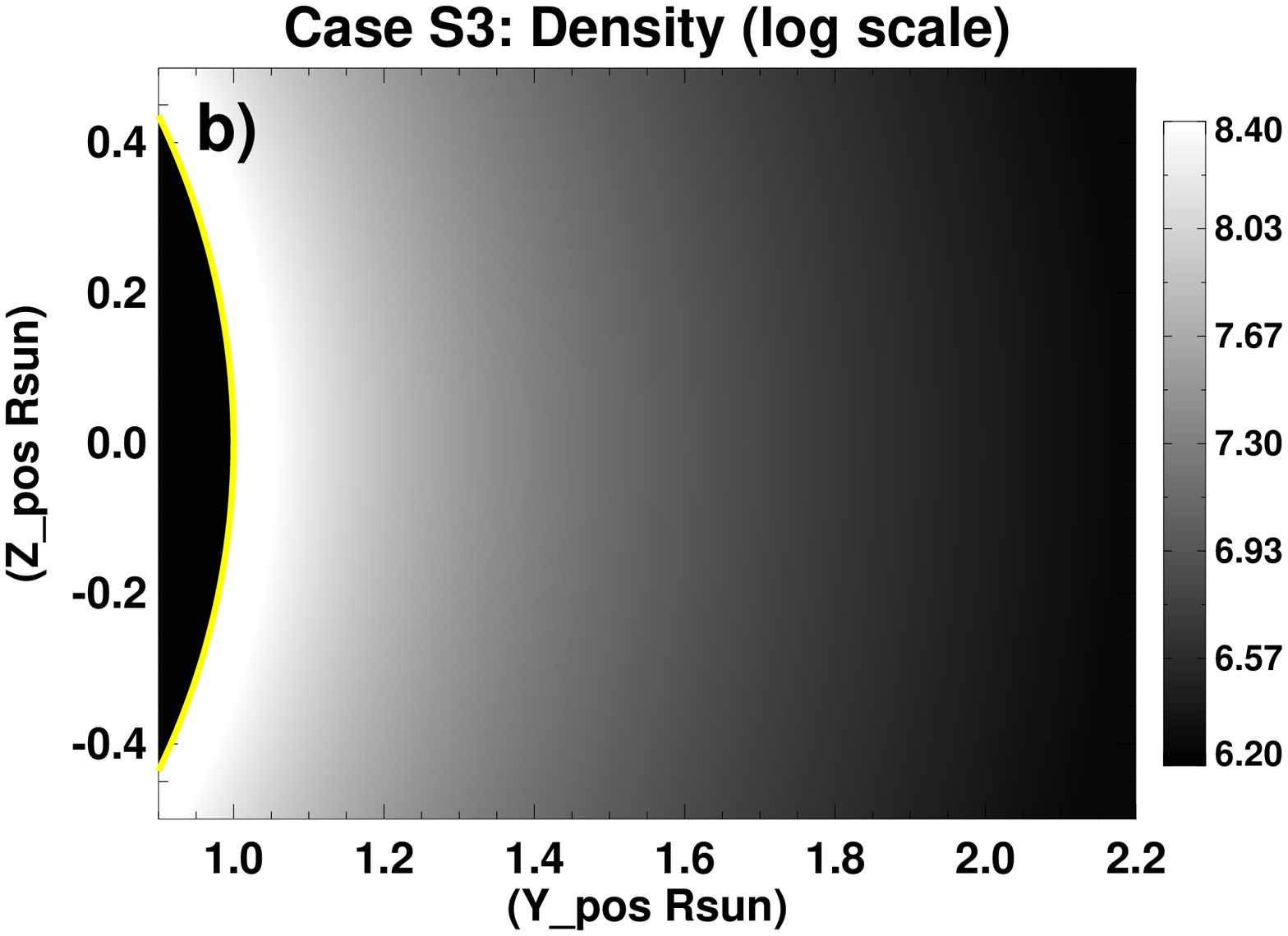}
              }
              \vspace{0.01\textwidth}
   \centerline{\hspace*{0.015\textwidth}
     \includegraphics[width=0.4\textwidth,clip=]{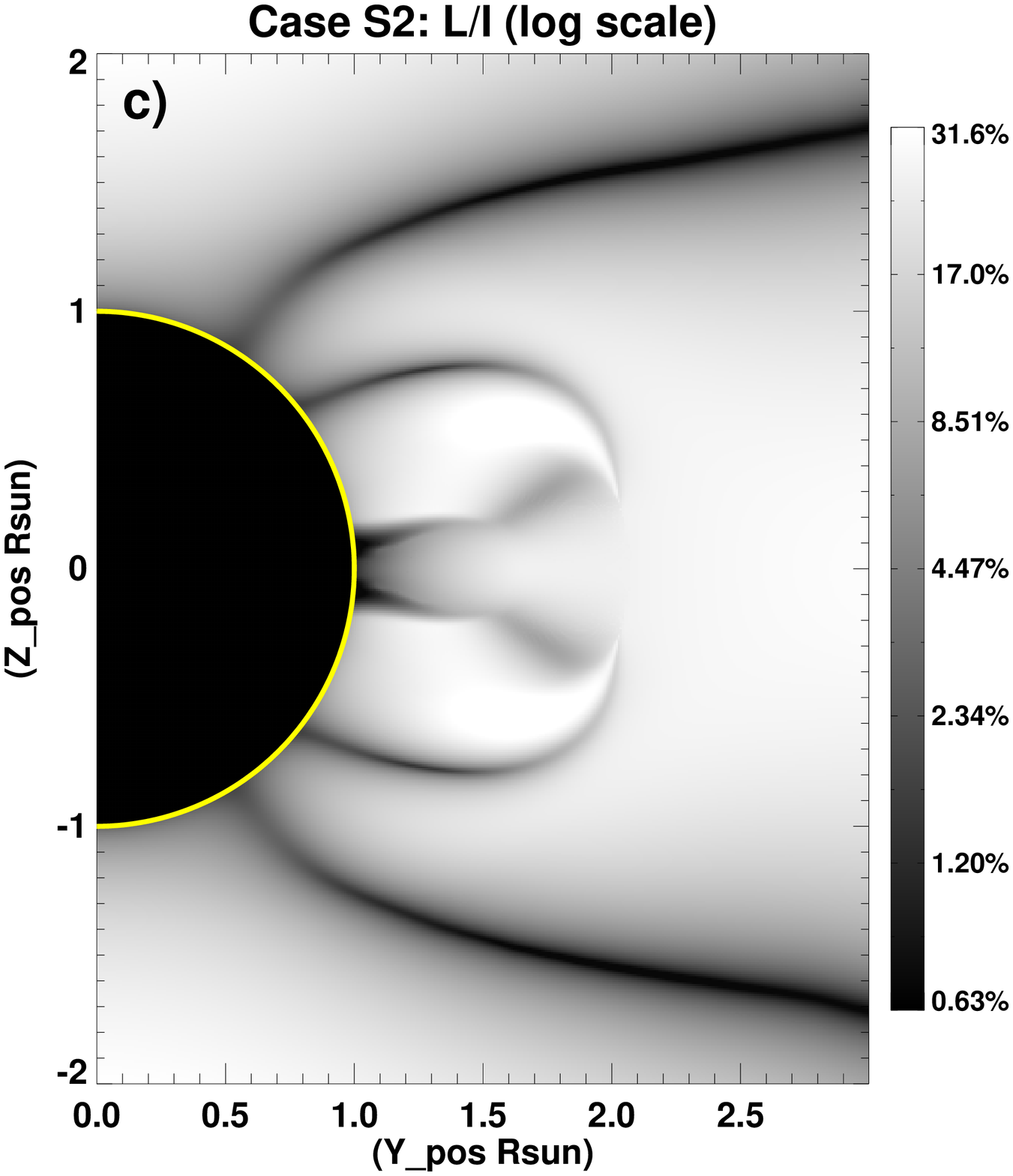}
               \includegraphics[width=0.4\textwidth,clip=]{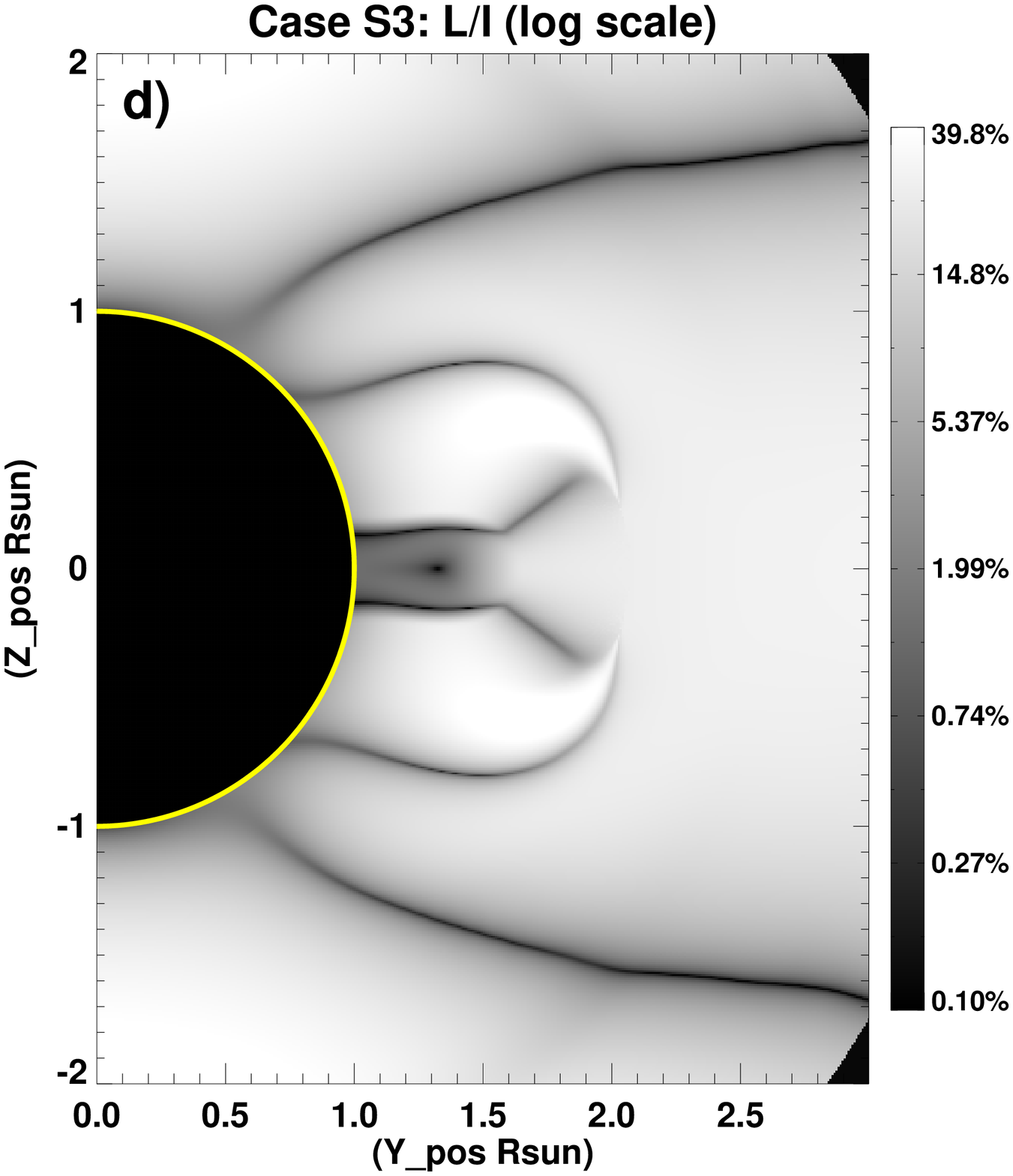}
              }
              \vspace{0.01\textwidth}
   \centerline{\hspace*{0.015\textwidth}
     \includegraphics[width=0.4\textwidth,clip=]{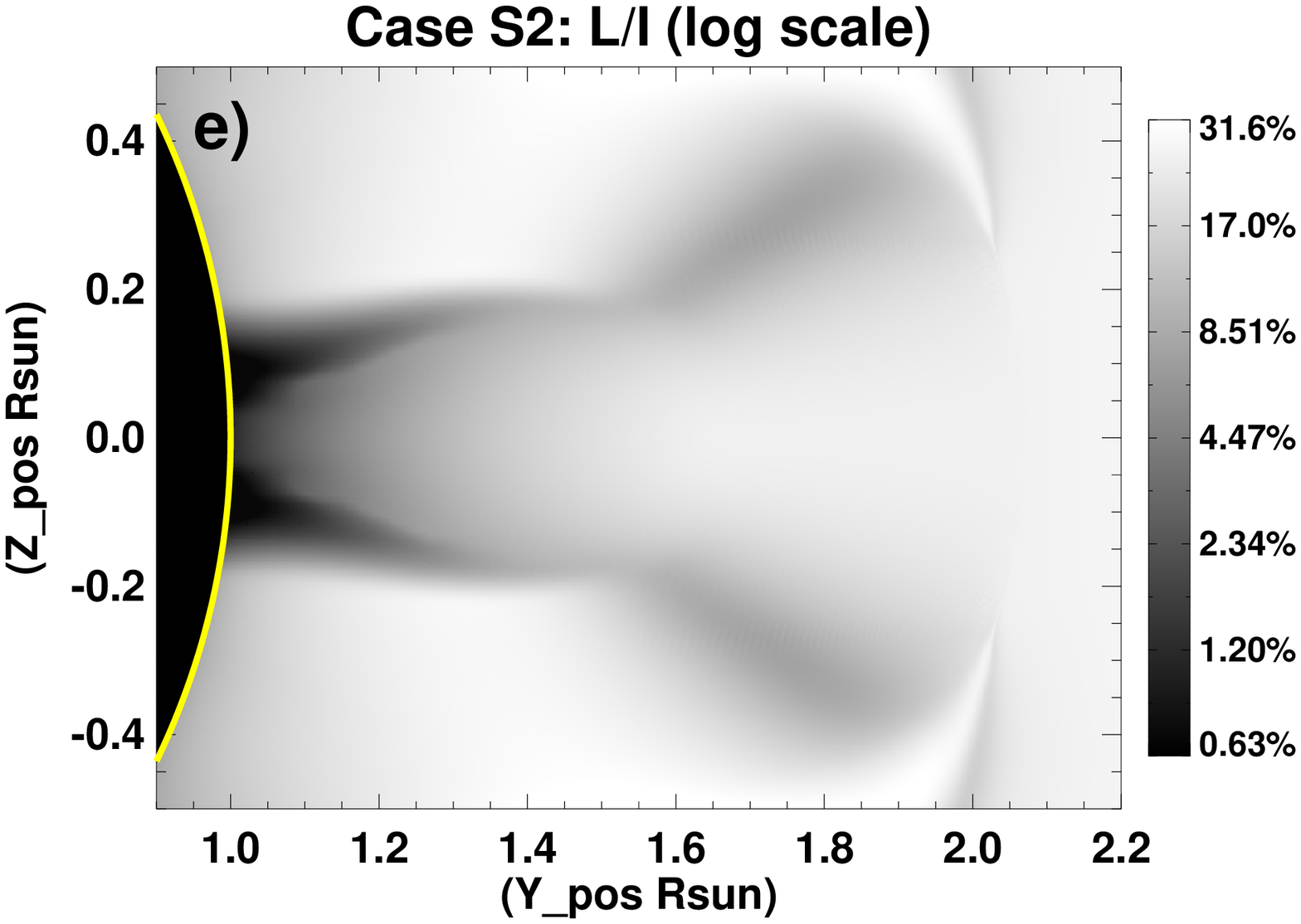}
               \includegraphics[width=0.4\textwidth,clip=]{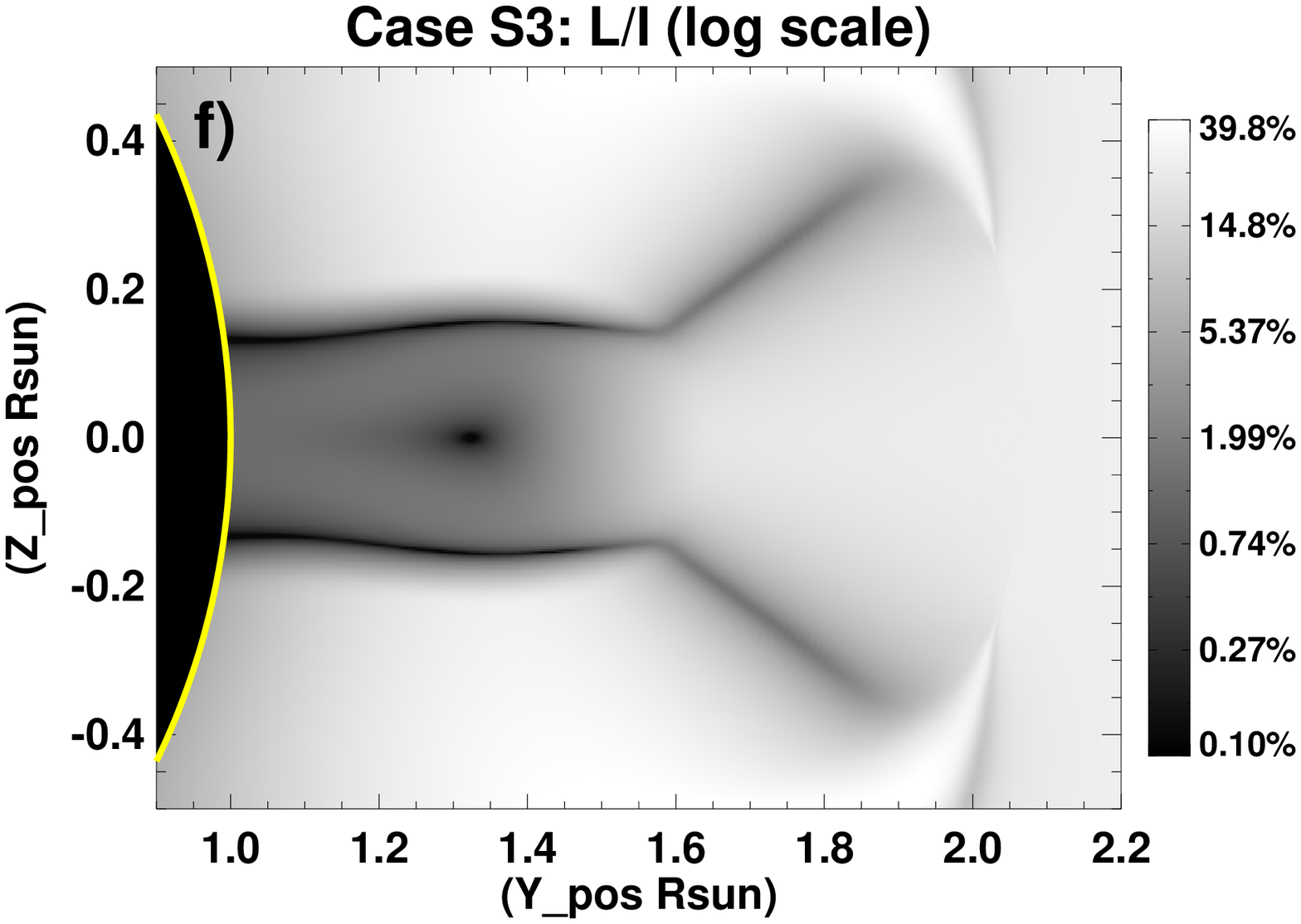}
              }
              \vspace{0.01\textwidth}
   \centerline{\hspace*{0.015\textwidth}
     \includegraphics[width=0.4\textwidth,clip=]{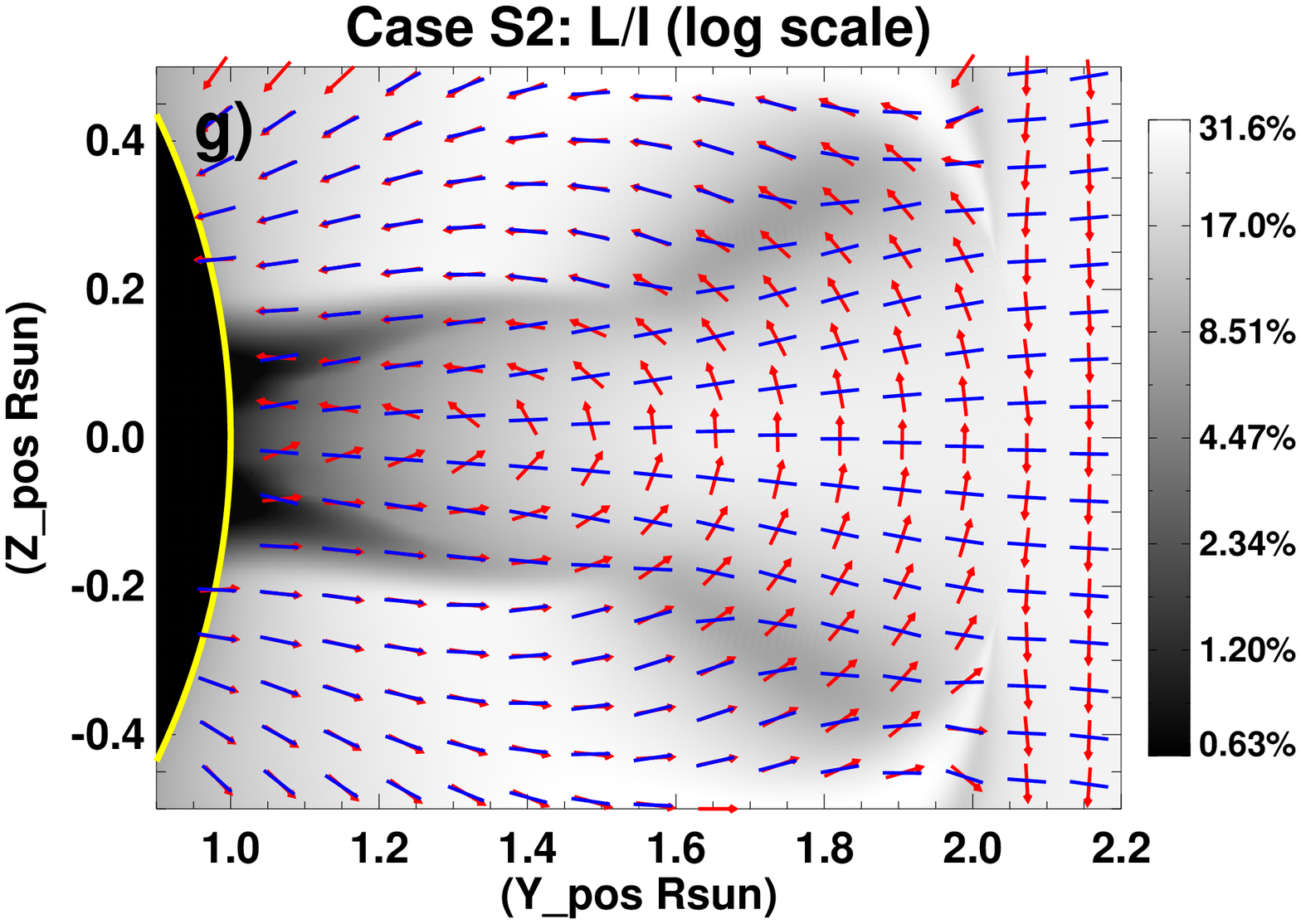}
               \includegraphics[width=0.4\textwidth,clip=]{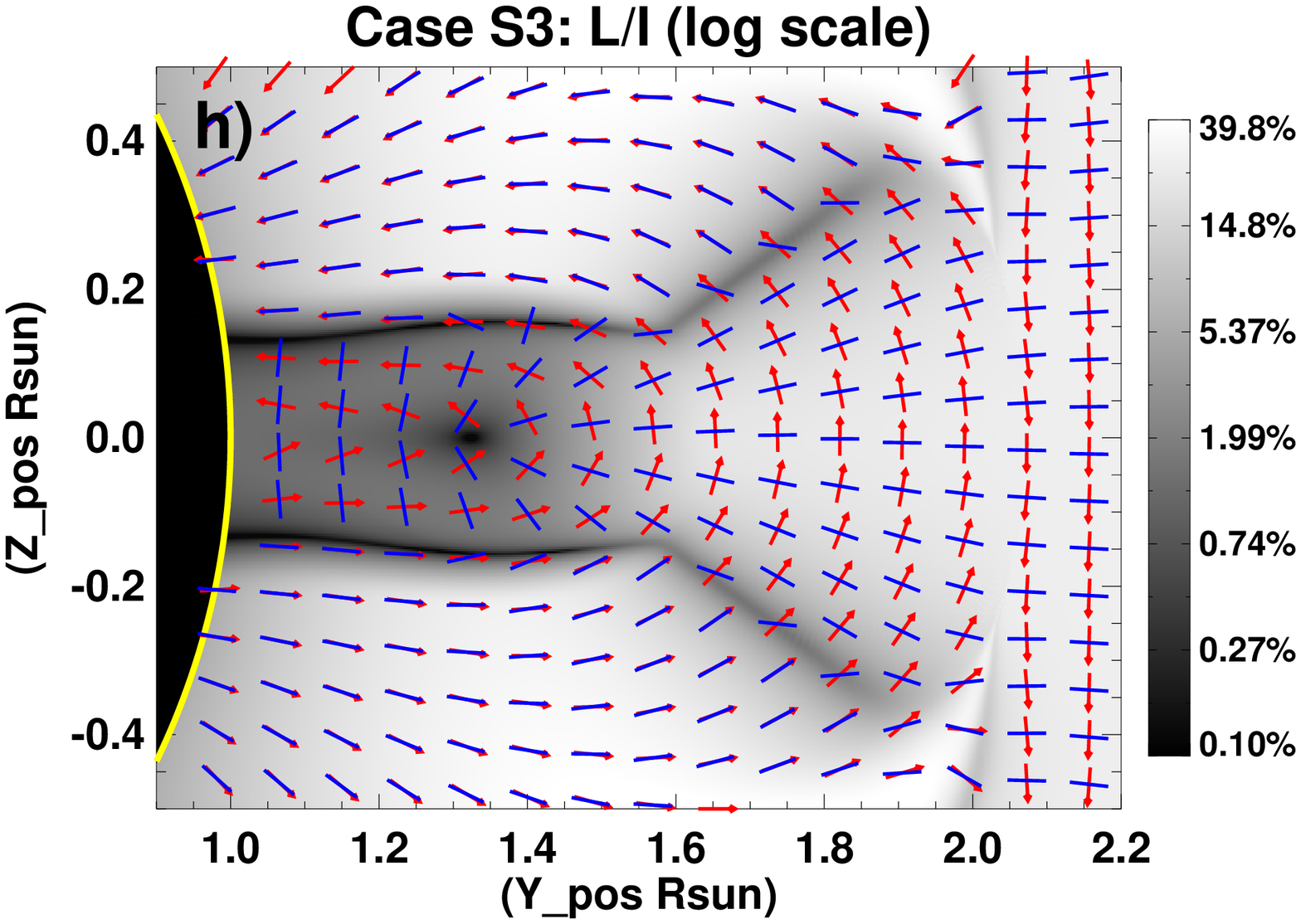}
              }
              \vspace{0.01\textwidth}
   \centerline{\hspace*{0.015\textwidth}
     \includegraphics[width=0.4\textwidth,clip=]{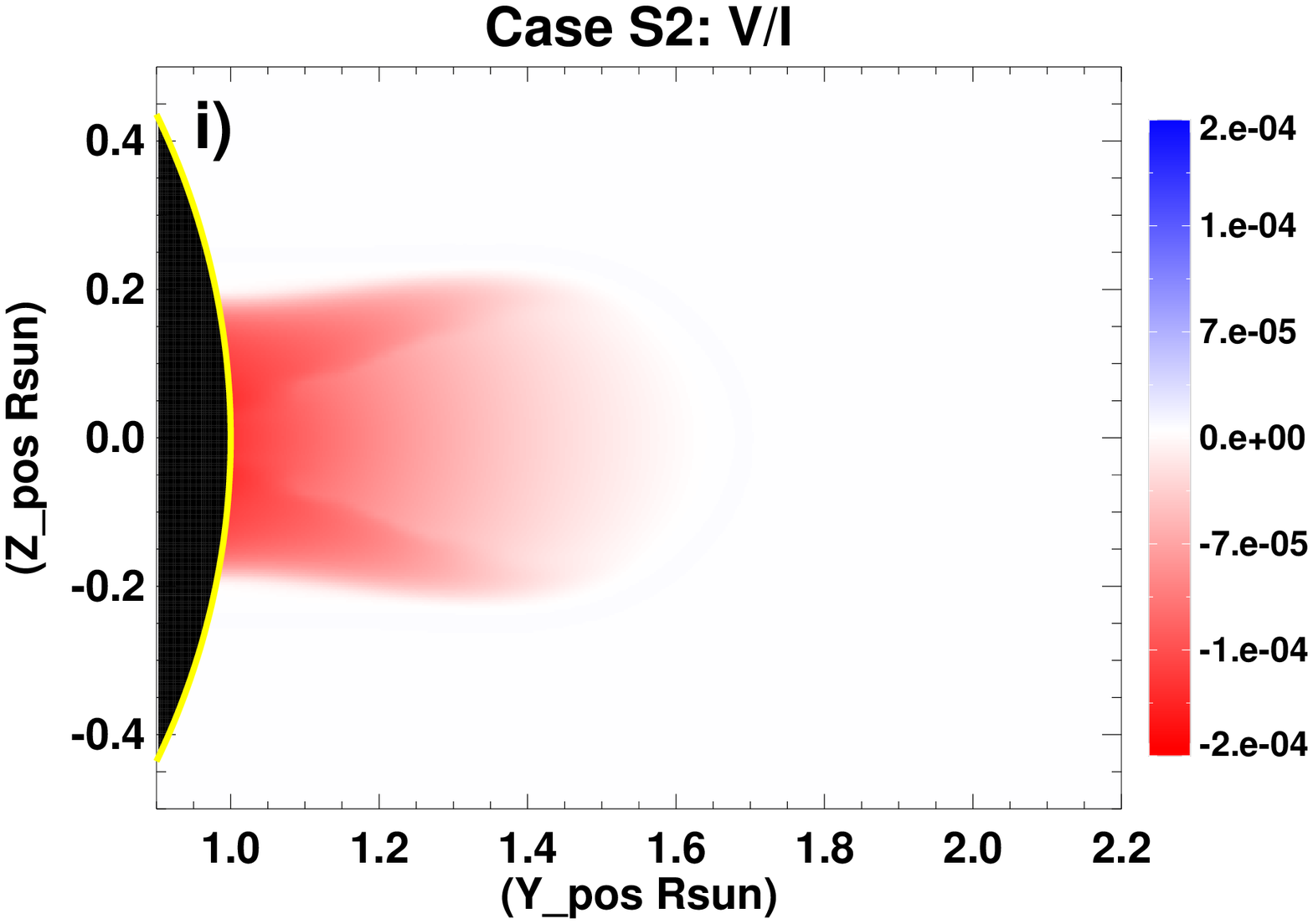}
               \includegraphics[width=0.4\textwidth,clip=]{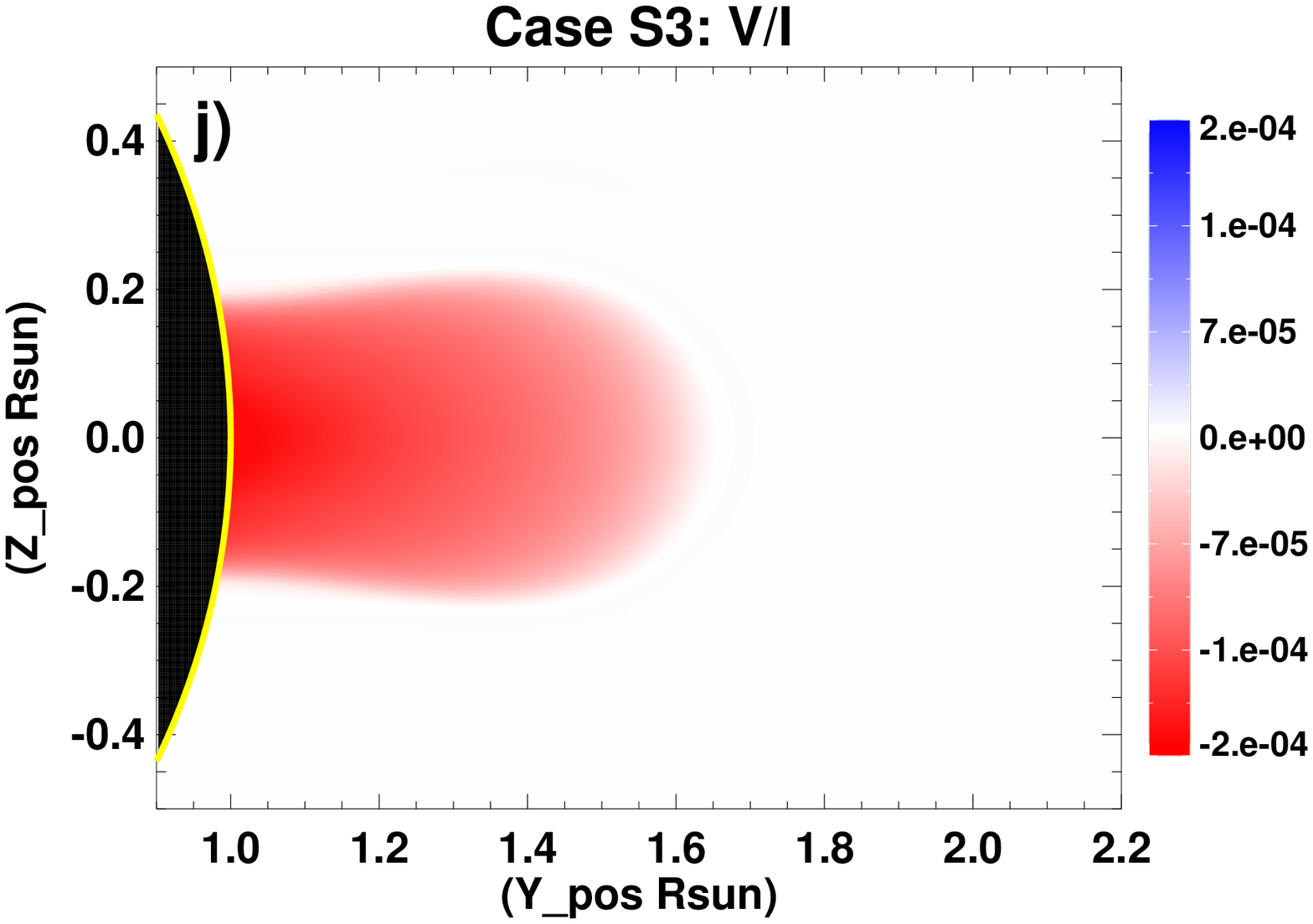}
              }
     
\caption{Forward-model results from the sheared arcade system. The left column
shows case S2 and the right column shows case S3. a) and b) density
in the central region, the electron number density is given by $10^{X}$~cm$^{-3}$ where $X$ is the value indicated by the color bar; c) and d) relative linear polarization of the
entire system; e) and f) relative linear polarization of the central
region; g) and h) same as above overlaid with red field vectors and
blue azimuth directions; i) and j) relative circular polarization
of the inner region. }
\label{fig:Devore-results}
\end{figure}

Figure \ref{fig:Devore-results}(g),(h) shows the comparison of the integrated
$L/I$ for the S2 and S3 cases in the sheared region. We have looked
at the polarization signals from thin POS slices along the LOS, and
find that for any given slice, the $L/I$ is qualitatively similar
between the two cases. The differences seen in Figure \ref{fig:Devore-results}
arise from the relative contributions to the integrated signal from
the sheared field versus the background unsheared field. Stokes $I$,
$Q$, and $U$ are weighted by density. Hence, higher-density regions
contribute more to the integrated signal than lower density regions.
In S2, the original density profile is used, and in S3 a spherically
symmetric density is used. These densities are plotted in Figure \ref{fig:Devore-results}(a),(b).
Note that in S2, the density in the unsheared field is about an order of magnitude 
more dense than the sheared field due to the large volumetric expansion 
of the sheared field and the closed lower boundary condition imposed in 
the simulation. From a broad range of observations,
it is known that streamers tend to be about a factor of two more dense
than the embedded cavity \cite{Fuller2009,Schmit2011}. Thus, the
S3 results put too much emphasis on the sheared field, and the S2
results put too little emphasis on this region compared to observations. 

\begin{figure}  

   \centerline{\hspace*{0.015\textwidth}
     \includegraphics[width=0.5\textwidth,clip=]{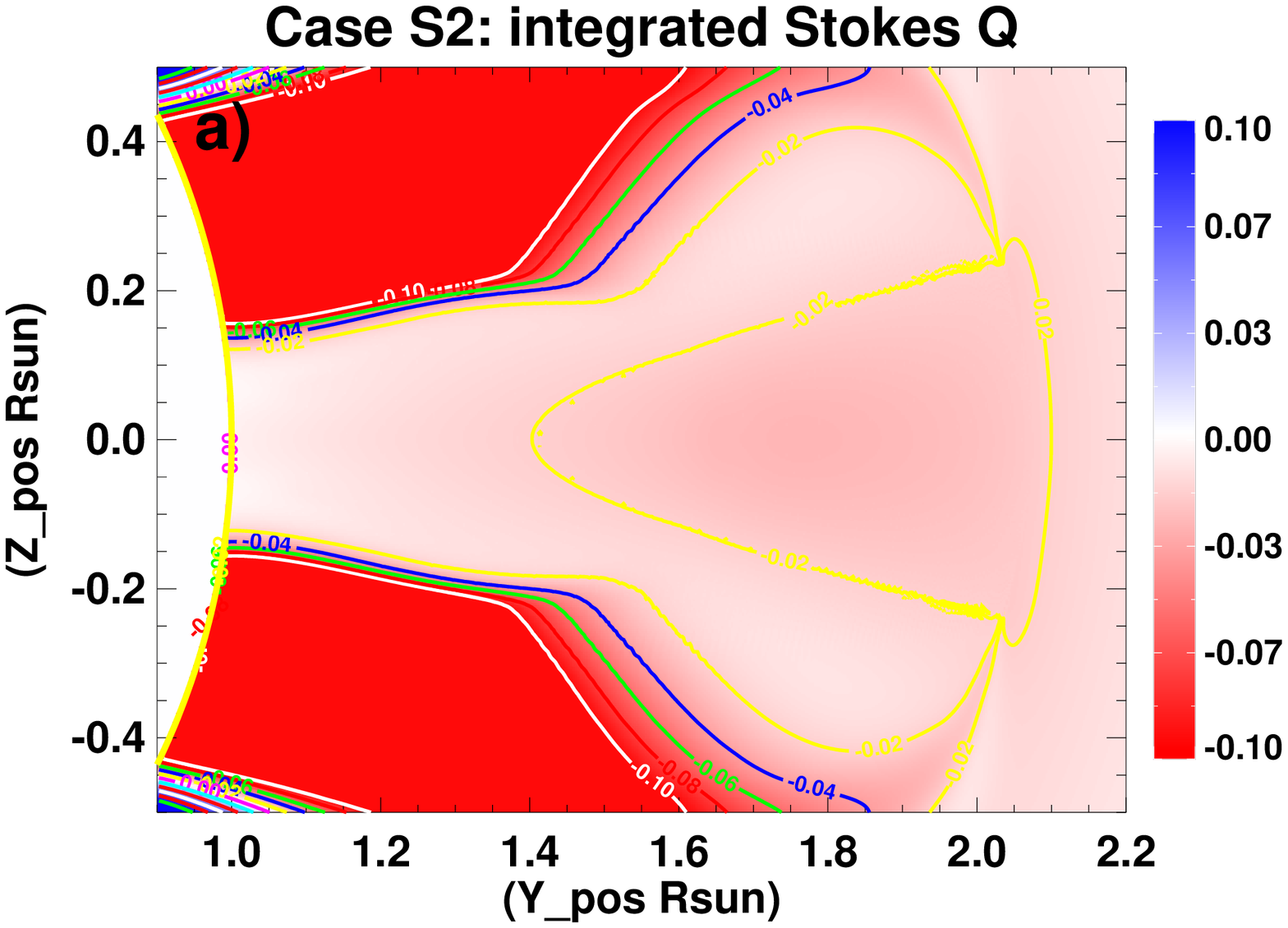}
               \includegraphics[width=0.5\textwidth,clip=]{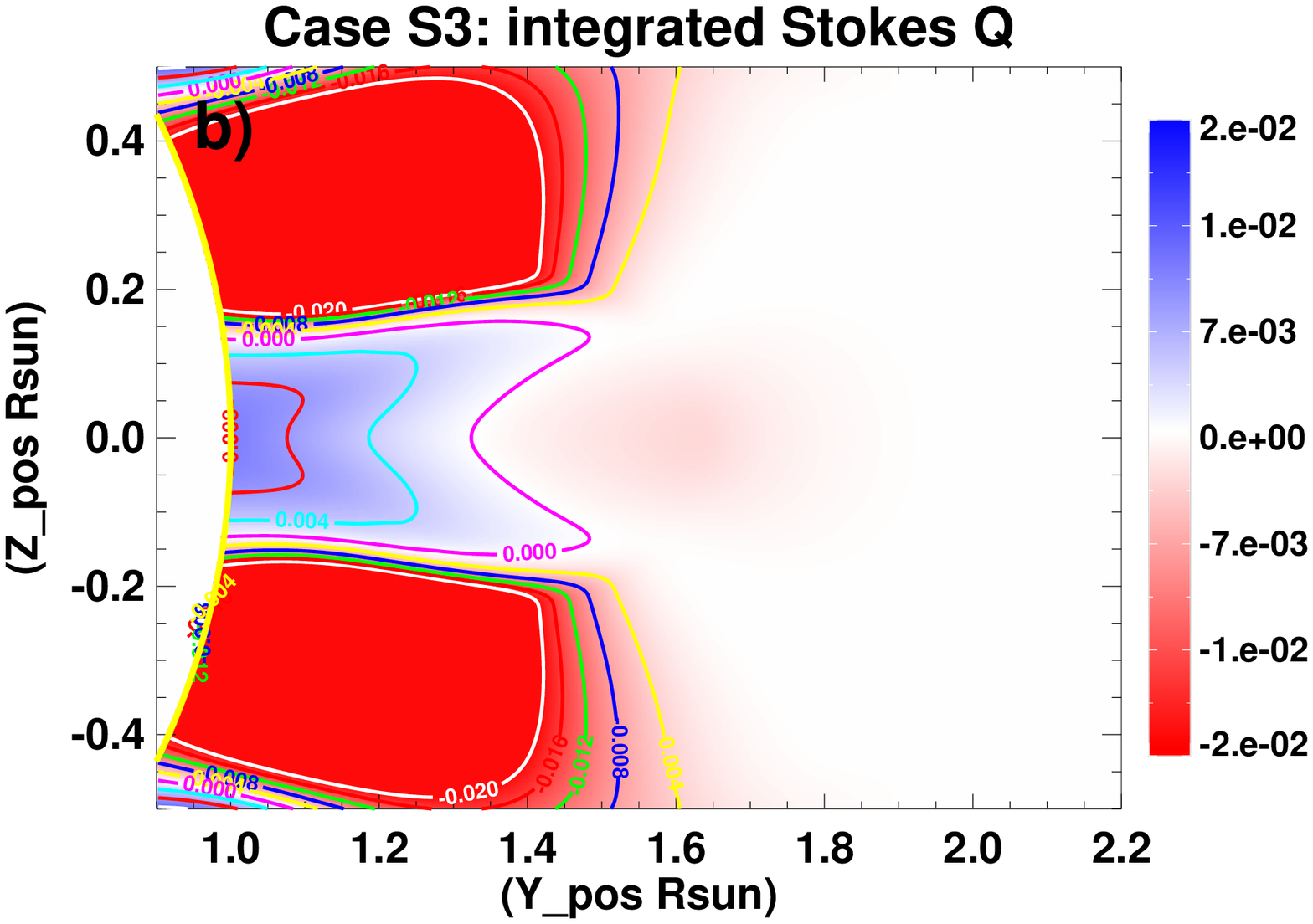}
              }
              \vspace{0.01\textwidth}
   \centerline{\hspace*{0.015\textwidth}
     \includegraphics[width=0.5\textwidth,clip=]{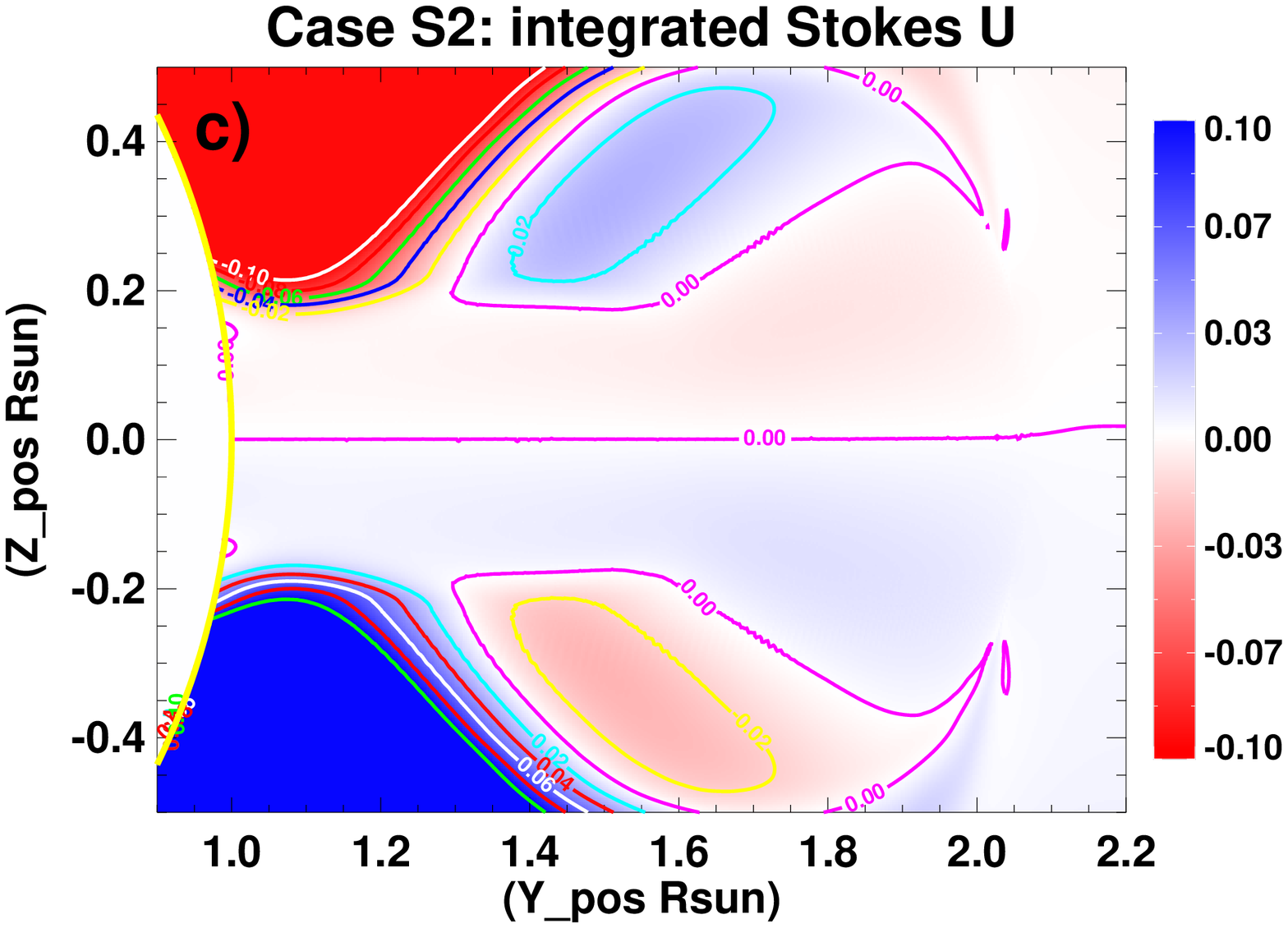}
               \includegraphics[width=0.5\textwidth,clip=]{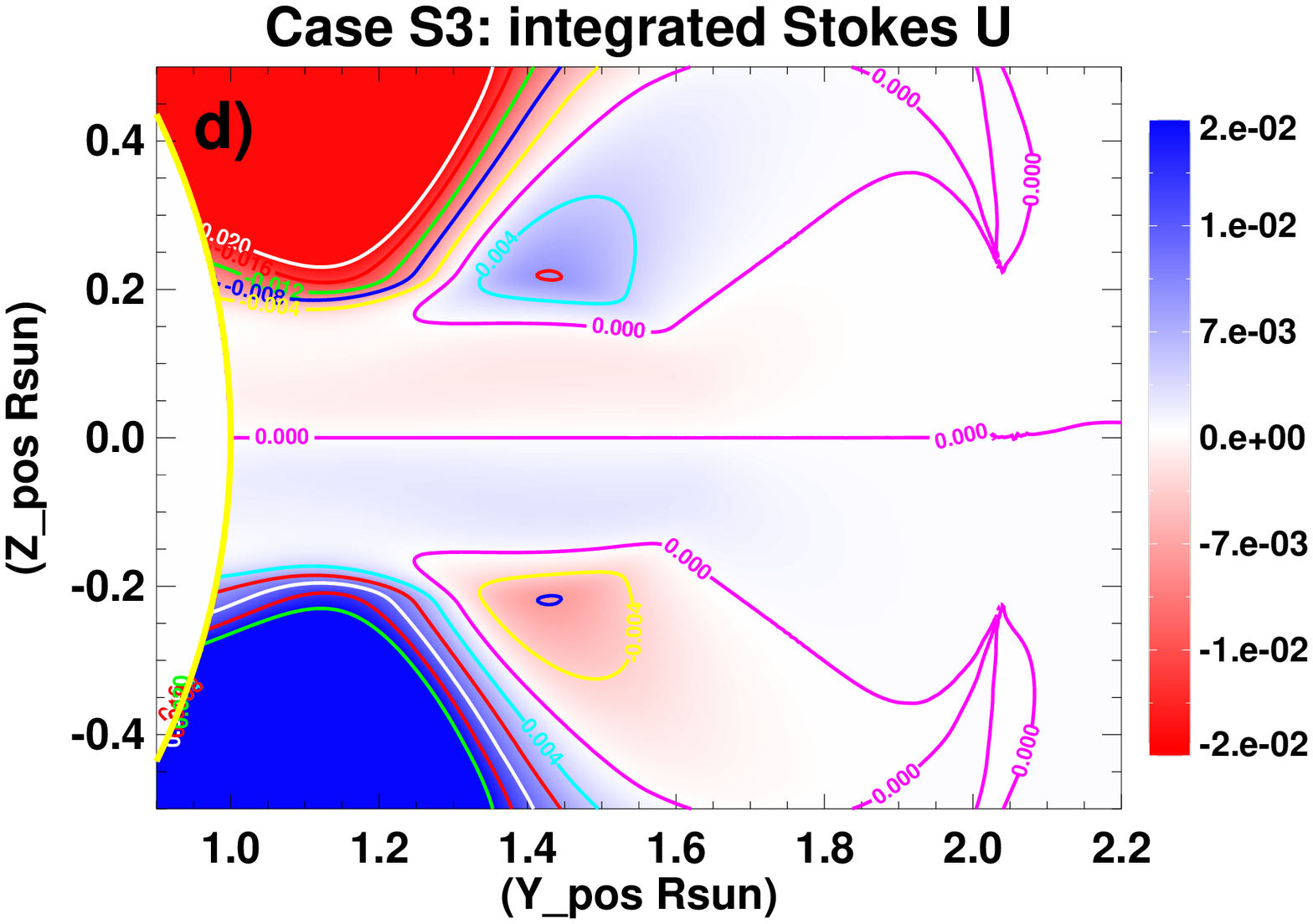}
              }
     
\caption{Integrated Stokes $Q$ and $U$ signals from cases S2 and S3 of the sheared arcade model.
Lines of zero $Q$ or $U$ are shown in magenta. }
\label{fig:devore-QU}
\end{figure}

The following list describes the features that are present in the data. Where certain features are found to a lesser degree, or not found at all, in one of the cases, it is explained in more detail after the list. 
\begin{enumerate}
	\item \emph{Quadrupolar Van Vleck signal}. A clear Van Vleck signature is
associated with the quadrupolar field (Figure~\ref{fig:Devore-results}(c),(d)).
Even in the absence of any sheared field, there would be three pairs
of Van Vleck inversions associated with the inner three loop systems.
These pairs are the top two, the middle two, and the lower two elongated Van
Vleck nulls. 
	\item \emph{$\vee$-shaped Van Vleck inversion.} At the top of the central loop system, there are Van Vleck lines in a $\vee$-shape
(Figure~\ref{fig:Devore-results}(e),(f); at a height of $1.6-2$~R$_{\odot}$), which is analogous to property
i) in the cylindrical flux-rope model. The sheared field is confined
to the central part of this system ({\it i.e.} the areas of negative $V/I$
in Figure~\ref{fig:Devore-results}(i),(j)) and the $\vee$-shaped Van
Vleck lines are associated with the unsheared portion of this central
magnetic-loop system (dark-blue loops above the sheared field in Figure
\ref{fig:devore-fieldlines}). 
	\item \emph{Parallel Van Vleck inversions.} In the sheared field,
there are two dark parallel Van Vleck inversion lines 
(Figure~\ref{fig:Devore-results}(e),(f); at a height of $1-1.6$~R$_{\odot}$)
that emanate from the photosphere
and connect to the $\vee$-shape listed in ii). The parallel Van Vleck
inversion lines are associated with the legs of the sheared region.
This field is inclined toward/away from the observer at $\approx54^{\circ}$
from radial. These inversions are more pronounced in S3 than in S2.
	\item \emph{Dark LOS core in L/I.} The central sheared area has generally
lower linear polarization magnitude due to the presence
of fields that are more LOS than the surrounding field. This effect is stronger in S3 than in S2.
	\item \emph{Anomalous LOS signal in L/I.} A dark spot in $L/I$ is visible
in the sheared field region in case S3 (Figure~\ref{fig:Devore-results}(f); $Z = 0$, $Y = 1.35$).
This is not due to an axis of LOS field. This anomalous LOS signal
\cite{Rachmeler2012} arises from cancellation in the LOS integrated
Stokes $Q$ and $U$ due to the symmetry in the system and is dependent
on the relative density in the sheared and unsheared regions. This
feature is not present in S2 (Figure~\ref{fig:Devore-results}(e)).
	\item \emph{Non-radial azimuth direction.} In the sheared-field region,
the linear polarization azimuth direction is parallel to the limb in the S3 case
(blue lines in Figure~\ref{fig:Devore-results}(h)), which creates an illusion of
a cylindrical flux rope. This is an area where the magnetic-field
angle exceeds the Van Vleck angle, so the true field is perpendicular
to the observed azimuth direction. Note that at this location, the
field is actually predominantly in the LOS, and the POS component is small. This feature is not present in S2 (Figure~\ref{fig:Devore-results}(g)).
	\item \emph{Strongest circular polarization near limb (V/I)}. The relative circular
polarization is shown in Figure~\ref{fig:Devore-results}(i),(j). The
shear is not concentrated above the limb, as in the flux-rope case,
but at the limb. The features within the negative $V/I$ in S2 are due to density variations (Figure \ref{fig:Devore-results}(a)). The $V/I$ in
the S3 case shows a smooth profile. 
\end{enumerate}
Most of these features are present in both cases, but they are not
always obvious in the integrated S2 data. The locations of the Van
Vleck inversion and the LOS field i)\,--\,iii) are the same
in both cases. Figure \ref{fig:devore-QU} shows the LOS integrated values 
of Stokes $Q$ and $U$ [$L=\sqrt{Q^{2}+U^{2}}$] for cases S2 and S3.
The locations of lowest $Q$ and $U$ are the same
for both cases, but in S2, there are no true nulls in $Q$ inside the sheared-field region. Thus,
the integrated $L/I$ for S2 (Figure~\ref{fig:Devore-results}(e)) does
not show true Van Vleck nulls. The inversions are there, but
the signal from the unsheared portion of the LOS obscures them.

There is clearly no anomalous LOS signal iv) in S2 (Figure~\ref{fig:Devore-results}(e)).
This is because the background dominates the integrated signal at
all heights in this case. Investigation of the polarization generated in thin POS
 slices shows that the unsheared field
in all cases produces a negative Stokes $Q$ signal, and the sheared
field produces signal that is predominantly positive. The relative
density in the sheared and background field dictates where the sheared
field dominates in the integrated Stokes profiles. The integrated
Stokes $Q$ signal for S3 (Figure~\ref{fig:devore-QU}(b)) is positive
in the central region, showing that the sheared field dominates the
LOS signal there. Where a zero-line in Stokes $Q$ crosses a
zero-line in Stokes $U$ (Figure~\ref{fig:devore-QU}(b),(d)) there is an
anomalous LOS signal (Figure~\ref{fig:Devore-results}(f))
\cite{Rachmeler2012}. Since the sheared field
in S2 has less density, the integrated Stokes $Q$ is always dominated
by the background field (Figure~\ref{fig:devore-QU}(a)) and is never
negative. Thus there is no anomalous LOS signal in S2. 

The relative weighting of the sheared versus the unsheared field
also causes the difference in the azimuth direction for cases S2 and
S3. The volume of space that contains sheared field lies inside a Van Vleck inversion -- such that the azimuth is perpendicular to the POS magnetic-field direction -- and inspections of a thin POS slice reveals azimuth directions that are consistent
with Figure~\ref{fig:Devore-results}(h). The integrated $L/I$ in case
S2 shows radial azimuths because the background
field dominates, so the non-radial azimuth signal is overwhelmed. 

\section{Discussion \label{sec:discussion}}

We have presented three coronal models and their synthetic polarization
signatures. We find that each of these models is distinct and distinguishable,
even when using linear polarization alone. The spheromak flux rope
is the most recognizably different, while the cylindrical flux rope
and the sheared arcade models have some similarity. 

This work highlights the importance of using a forward approach on
coronal emission-line polarization. It can teach us what to look for
in observations, such as Van Vleck inversions and LOS field. It also calls
attention to the fact that we cannot trust our intuition to pick out
magnetic morphologies. The sheared-arcade model is a good example of
this. On initial inspection, the polarization signatures of the S3
sheared arcade resemble a flux rope; the azimuth direction is parallel
to the limb between the inner Van Vleck inversions, and a false axial
signature may be present. Both of these signatures can be logically
explained when analyzing the forward results, but this example underlines
the need for forward or inverse analysis before magnetic-structure
identification can be made. Another important strength of the forward
approach is that it fully takes into account the LOS integration of
the polarization signal. The presence of the optically thin plasma
is a significant challenge for the inverse technique and so is often
seen as a limitation for coronal polarization data as a whole, but
the forward approach incorporates the lack of opacity.
By looking at a given magnetic configuration with multiple plasma
profiles, we can also learn about how the signatures change
due to the plasma parameters alone. We have shown that in the case
of cavities, where the cavity itself is about half as dense as the
surrounding streamer, the polarization signature from the streamer
can obscure some of the features that are present in the cavity.
For future observations, it is clear that knowledge of the LOS density
structures is important for analysis and interpretation. Density diagnostics
which determine a 3D density distribution could be used in conjunction
with polarization observations and forward calculations.

Our work is not the first to use this forward approach to understand
hypothetical or actual observations. \inlinecite{Judge2006} applied the
same forward code to study prominence-supporting magnetic fields and
current sheets. \inlinecite{Liu2008} compared observations of an active
region on the limb to potential-field extrapolations to study how
the LOS affects the fit of the forward calculation to the observations.
\inlinecite{Dove2011} compared the spheromak model presented here to an
early CoMP observation of a large cavity. The next important step
is to take the knowledge gained with these forward studies and apply
it directly to observations, looking for the specific morphologies; this work is already underway.
\inlinecite{BakSteslicka2013} have found that cavities
observed in CoMP in 2011 and 2012 usually have a characteristic ``rabbit-head''
signature in $L/I$. This signature consists of two Van Vleck inversions, with or without a dark
central region indicating LOS field. They have shown that this observation
is consistent with a 3D flux-rope topology where the height of the
dark central ``head'' is approximately co-spatial with the center of
the cavity. 

Observations carry their own challenges because of noise and the existence
of small-scale density structures in the corona. CoMP is an occulted
coronagraph and the occulter is at approximately $1.05$~ R$_{\odot}$,
which means that especially for small cavities, the distinguishing characteristics
such as azimuth direction would likely be obscured by the occulter.
We have used extremely simplified density structures to isolate the
magnetic features, but in reality, the Stokes $I$ observations are
highly structured. By analyzing \emph{relative} linear and circular
polarization, we remove some of the density component, but the signal
is still density dependent and we have shown that the relative importance
of the structures along the LOS is highly dependent on their density. Our current approach is more applicable in coronal cavities
which are, in general, fairly smooth in intensity compared to active
regions with clear bright loops. In future forward-modeling
research, more realistic density models are needed. The observational
noise, the occulter, and the highly structured coronal density make
it difficult to uniquely characterize observed cavities, as they may
be consistent with more than one model. 

We are just beginning to scratch the surface of what the polarization
data can teach us about the solar corona. Here we have studied idealized
equilibrium structures. Not only is there a large range of magnetic
morphologies left to study, there is also the important aspect of
time-dependence that is still open for exploration. The forward approach
is only one of the methods available for analyzing these data, and
there is still much to do with comparisons to observations, true forward
fitting for given observations, and looking at the Sun as a whole
as opposed to specific magnetic structures. We look forward to witnessing
the advances that come out of these data in conjunction with both forward
and inverse techniques. 
\comment{
STILL NEED:
-real reference to Bak-Steslicka2013 once it gets published
}

%
 \begin{acks}
The authors would like to thank R. Casini, P. Judge, and S. Tomczyk for many helpful conversations on coronal polarization and CoMP in general, and this article in specific. We would also like to thank all of those who have contributed to the continued expansion of the {\sf FORWARD} model. LAR acknowledges support from HAO visitors' fund, STFC (UK), and PRODEX grant C90193 managed by the European Space Agency in collaboration with the Belgian Federal Science Policy Office for financial support. This work was also supported in part by NASA grant NNX08AU30G. CRD acknowledges support from NASA for his participation. NCAR is sponsored by the National Science Foundation.
 \end{acks}

%
 \bibliographystyle{spr-mp-sola}
 \bibliography{stokes}  

\begin{thebibliography}{48}
\ifx \bisbn   \undefined \def \bisbn  #1{ISBN #1}\fi
\ifx \binits  \undefined \def \binits#1{#1}\fi
\ifx \bauthor  \undefined \def \bauthor#1{#1}\fi
\ifx \batitle  \undefined \def \batitle#1{#1}\fi
\ifx \bjtitle  \undefined \def \bjtitle#1{\textit{#1}}\fi
\ifx \bvolume  \undefined \def \bvolume#1{\textbf{#1}}\fi
\ifx \byear  \undefined \def \byear#1{#1}\fi
\ifx \bissue  \undefined \def \bissue#1{#1}\fi
\ifx \bfpage  \undefined \def \bfpage#1{#1}\fi
\ifx \blpage  \undefined \def \blpage #1{#1}\fi
\ifx \burl  \undefined \def \burl#1{\textsf{#1}}\fi
\ifx \href  \undefined \def \href#1#2{\textsf{#2}}\fi
\ifx \doiurl  \undefined \def
  \doiurl#1{\href{http://dx.doi.org/#1}{\textsf{#1}}}\fi
\ifx \betal  \undefined \def \betal{\textit{et al.}}\fi
\ifx \binstitute  \undefined \def \binstitute#1{#1}\fi
\ifx \bctitle  \undefined \def \bctitle#1{#1}\fi
\ifx \beditor  \undefined \def \beditor#1{#1}\fi
\ifx \bpublisher  \undefined \def \bpublisher#1{#1}\fi
\ifx \bbtitle  \undefined \def \bbtitle#1{\textit{#1}}\fi
\ifx \bedition  \undefined \def \bedition#1{#1}\fi
\ifx \bseriesno  \undefined \def \bseriesno#1{\textbf{#1}}\fi
\ifx \blocation  \undefined \def \blocation#1{#1}\fi
\ifx \bsertitle  \undefined \def \bsertitle#1{\textit{#1}}\fi
\ifx \bsnm \undefined \def \bsnm#1{#1}\fi
\ifx \bsuffix \undefined \def \bsuffix#1{#1}\fi
\ifx \bparticle \undefined \def \bparticle#1{#1}\fi
\ifx \barticle \undefined \def \barticle#1{}\fi
\ifx \botherref \undefined \def \botherref#1{}\fi
\ifx \url \undefined \def \url#1{\textsf{#1}}\fi
\ifx \bchapter \undefined \def \bchapter#1{}\fi
\ifx \bbook \undefined \def \bbook#1{}\fi
\ifx \bcomment \undefined \def \bcomment#1{#1}\fi
\ifx \oauthor \undefined \def \oauthor#1{#1}\fi
\ifx \citeauthoryear \undefined \def \citeauthoryear#1{#1}\fi
\def \endbibitem {}
\ifx \bconflocation  \undefined \def \bconflocation#1{#1} \fi

\bibitem[\protect\citeauthoryear{{Antiochos}, {Dahlburg}, and
  {Klimchuk}}{1994}]{Antiochos1994}
\begin{barticle}
\bauthor{\bsnm{{Antiochos}}, \binits{S.K.}},
\bauthor{\bsnm{{Dahlburg}}, \binits{R.B.}},
\bauthor{\bsnm{{Klimchuk}}, \binits{J.A.}}:
\byear{1994},
\batitle{{The magnetic field of solar prominences}}.
\bjtitle{\apjl}
\bvolume{420},
\bfpage{L41}\,--\,\blpage{L44}.
doi:\doiurl{10.1086/187158}.
\end{barticle}
\endbibitem

\bibitem[\protect\citeauthoryear{{Antiochos}, {DeVore}, and
  {Klimchuk}}{1999}]{Antiochos1999}
\begin{barticle}
\bauthor{\bsnm{{Antiochos}}, \binits{S.K.}},
\bauthor{\bsnm{{DeVore}}, \binits{C.R.}},
\bauthor{\bsnm{{Klimchuk}}, \binits{J.A.}}:
\byear{1999},
\batitle{{A Model for Solar Coronal Mass Ejections}}.
\bjtitle{\apj}
\bvolume{510},
\bfpage{485}\,--\,\blpage{493}.
doi:\doiurl{10.1086/306563}.
\end{barticle}
\endbibitem

\bibitem[\protect\citeauthoryear{{Arnaud} and {Newkirk}}{1987}]{Arnaud1987}
\begin{barticle}
\bauthor{\bsnm{{Arnaud}}, \binits{J.}},
\bauthor{\bsnm{{Newkirk}}, \binits{G.} \bsuffix{Jr.}}:
\byear{1987},
\batitle{{Mean properties of the polarization of the Fe XIII 10747 A coronal
  emission line}}.
\bjtitle{\aap}
\bvolume{178},
\bfpage{263}\,--\,\blpage{268}.
\end{barticle}
\endbibitem

\bibitem[\protect\citeauthoryear{{Aschwanden}
  \textit{et~al.}}{1999}]{Aschwanden1999}
\begin{barticle}
\bauthor{\bsnm{{Aschwanden}}, \binits{M.J.}},
\bauthor{\bsnm{{Newmark}}, \binits{J.S.}},
\bauthor{\bsnm{{Delaboudini{\`e}re}}, \binits{J.-P.}},
\bauthor{\bsnm{{Neupert}}, \binits{W.M.}},
\bauthor{\bsnm{{Klimchuk}}, \binits{J.A.}},
\bauthor{\bsnm{{Gary}}, \binits{G.A.}},
\bauthor{\bsnm{{Portier-Fozzani}}, \binits{F.}},
\bauthor{\bsnm{{Zucker}}, \binits{A.}}:
\byear{1999},
\batitle{{Three-dimensional Stereoscopic Analysis of Solar Active Region Loops.
  I. SOHO/EIT Observations at Temperatures of (1.0-1.5) X 10\^{}6 K}}.
\bjtitle{\apj}
\bvolume{515},
\bfpage{842}\,--\,\blpage{867}.
doi:\doiurl{10.1086/307036}.
\end{barticle}
\endbibitem

\bibitem[\protect\citeauthoryear{{Ba\c k-St\c e\' slicka}
  \textit{et~al.}}{2013}]{BakSteslicka2013}
\begin{botherref}
\oauthor{\bsnm{{Ba\c k-St\c e\' slicka}}, \binits{U.}},
\oauthor{\bsnm{{Gibson}}, \binits{S.E.}},
\oauthor{\bsnm{{Fan}}, \binits{Y.E.}},
\oauthor{\bsnm{{Bethge}}, \binits{C.W.}},
\oauthor{\bsnm{{Forland}}, \binits{B.}},
\oauthor{\bsnm{{Rachmeler}}, \binits{L.A.}}:
2013,
{The Magnetic Structure of Solar Prominence Cavities: New Observable}.
\textit{\apjl},
submitted.
\end{botherref}
\endbibitem

\bibitem[\protect\citeauthoryear{{Bastian}}{2005}]{Bastian2004}
\begin{bchapter}
\bauthor{\bsnm{{Bastian}}, \binits{T.S.}}:
\byear{2005},
\bctitle{{The Frequency Agile Solar Radiotelescope}}.
In: \beditor{\bsnm{{Gary}}, \binits{D.E.}},
\beditor{\bsnm{{Keller}}, \binits{C.U.}} (eds.)
\bbtitle{Solar and Space Weather Radiophysics},
\bsertitle{Astrophys and Space Sci Library}
\bseriesno{314},
\bfpage{47}\,--\,\blpage{69}.
doi:\doiurl{10.1007/1-4020-2814-8\_3}.
\end{bchapter}
\endbibitem

\bibitem[\protect\citeauthoryear{{Casini}}{2002}]{Casini2002}
\begin{barticle}
\bauthor{\bsnm{{Casini}}, \binits{R.}}:
\byear{2002},
\batitle{{The Hanle Effect of the Two-Level Atom in the Weak-Field
  Approximation}}.
\bjtitle{\apj}
\bvolume{568},
\bfpage{1056}\,--\,\blpage{1065}.
doi:\doiurl{10.1086/338986}.
\end{barticle}
\endbibitem

\bibitem[\protect\citeauthoryear{{Casini} and {Judge}}{1999}]{Casini1999}
\begin{barticle}
\bauthor{\bsnm{{Casini}}, \binits{R.}},
\bauthor{\bsnm{{Judge}}, \binits{P.G.}}:
\byear{1999},
\batitle{{Spectral Lines for Polarization Measurements of the Coronal Magnetic
  Field. II. Consistent Treatment of the Stokes Vector for Magnetic-Dipole
  Transitions}}.
\bjtitle{\apj}
\bvolume{522},
\bfpage{524}\,--\,\blpage{539}.
doi:\doiurl{10.1086/307629}.
\end{barticle}
\endbibitem

\bibitem[\protect\citeauthoryear{{Charvin}}{1965}]{Charvin1965}
\begin{barticle}
\bauthor{\bsnm{{Charvin}}, \binits{P.}}:
\byear{1965},
\batitle{{{\'E}tude de la polarisation des raies interdites de la couronne
  solaire. Application au cas de la raie verte {$\lambda$} 5303}}.
\bjtitle{Annal d'Astrophys}
\bvolume{28},
\bfpage{877}.
\end{barticle}
\endbibitem

\bibitem[\protect\citeauthoryear{{Dove} \textit{et~al.}}{2011}]{Dove2011}
\begin{barticle}
\bauthor{\bsnm{{Dove}}, \binits{J.B.}},
\bauthor{\bsnm{{Gibson}}, \binits{S.E.}},
\bauthor{\bsnm{{Rachmeler}}, \binits{L.A.}},
\bauthor{\bsnm{{Tomczyk}}, \binits{S.}},
\bauthor{\bsnm{{Judge}}, \binits{P.}}:
\byear{2011},
\batitle{{A Ring of Polarized Light: Evidence for Twisted Coronal Magnetism in
  Cavities}}.
\bjtitle{\apjl}
\bvolume{731},
\bfpage{L1+}.
doi:\doiurl{10.1088/2041-8205/731/1/L1}.
\end{barticle}
\endbibitem

\bibitem[\protect\citeauthoryear{{Fan} and {Gibson}}{2006}]{fan200_yfan2d1}
\begin{barticle}
\bauthor{\bsnm{{Fan}}, \binits{Y.}},
\bauthor{\bsnm{{Gibson}}, \binits{S.E.}}:
\byear{2006},
\batitle{{On the Nature of the X-Ray Bright Core in a Stable Filament
  Channel}}.
\bjtitle{\apjl}
\bvolume{641},
\bfpage{L149}\,--\,\blpage{L152}.
doi:\doiurl{10.1086/504107}.
\end{barticle}
\endbibitem

\bibitem[\protect\citeauthoryear{{Fuller} and {Gibson}}{2009}]{Fuller2009}
\begin{barticle}
\bauthor{\bsnm{{Fuller}}, \binits{J.}},
\bauthor{\bsnm{{Gibson}}, \binits{S.E.}}:
\byear{2009},
\batitle{{A Survey of Coronal Cavity Density Profiles}}.
\bjtitle{\apj}
\bvolume{700},
\bfpage{1205}\,--\,\blpage{1215}.
doi:\doiurl{10.1088/0004-637X/700/2/1205}.
\end{barticle}
\endbibitem

\bibitem[\protect\citeauthoryear{{Gelfreikh}}{1994}]{Gelfreikh1994}
\begin{bchapter}
\bauthor{\bsnm{{Gelfreikh}}, \binits{G.B.}}:
\byear{1994},
\bctitle{{Radio measurements of coronal magnetic fields.}}
In: \beditor{\bsnm{{Rusin}}, \binits{V.}},
\beditor{\bsnm{{Heinzel}}, \binits{P.}},
\beditor{\bsnm{{Vial}}, \binits{J.-C.}} (eds.)
\bbtitle{Solar Coronal Structures},
\bsertitle{IAU Colloquium},
\bfpage{21}\,--\,\blpage{28}.
\end{bchapter}
\endbibitem

\bibitem[\protect\citeauthoryear{{Gibson} and {Low}}{1998}]{Gibson1998_giblow}
\begin{barticle}
\bauthor{\bsnm{{Gibson}}, \binits{S.E.}},
\bauthor{\bsnm{{Low}}, \binits{B.C.}}:
\byear{1998},
\batitle{{A Time-Dependent Three-Dimensional Magnetohydrodynamic Model of the
  Coronal Mass Ejection}}.
\bjtitle{\apj}
\bvolume{493},
\bfpage{460}.
doi:\doiurl{10.1086/305107}.
\end{barticle}
\endbibitem

\bibitem[\protect\citeauthoryear{{Gibson} and {Low}}{2000}]{gibson2000_giblow}
\begin{barticle}
\bauthor{\bsnm{{Gibson}}, \binits{S.E.}},
\bauthor{\bsnm{{Low}}, \binits{B.C.}}:
\byear{2000},
\batitle{{Three-dimensional and twisted: An MHD interpretation of on-disk
  observational characteristics of coronal mass ejections}}.
\bjtitle{\jgr}
\bvolume{105},
\bfpage{18187}\,--\,\blpage{18202}.
doi:\doiurl{10.1029/1999JA000317}.
\end{barticle}
\endbibitem

\bibitem[\protect\citeauthoryear{{Gibson} \textit{et~al.}}{1999}]{Gibson1999}
\begin{barticle}
\bauthor{\bsnm{{Gibson}}, \binits{S.E.}},
\bauthor{\bsnm{{Fludra}}, \binits{A.}},
\bauthor{\bsnm{{Bagenal}}, \binits{F.}},
\bauthor{\bsnm{{Biesecker}}, \binits{D.}},
\bauthor{\bsnm{{del Zanna}}, \binits{G.}},
\bauthor{\bsnm{{Bromage}}, \binits{B.}}:
\byear{1999},
\batitle{{Solar minimum streamer densities and temperatures using Whole Sun
  Month coordinated data sets}}.
\bjtitle{\jgr}
\bvolume{104},
\bfpage{9691}\,--\,\blpage{9700}.
doi:\doiurl{10.1029/98JA02681}.
\end{barticle}
\endbibitem

\bibitem[\protect\citeauthoryear{{Gibson}
  \textit{et~al.}}{2006}]{Gibson2006_cavities}
\begin{barticle}
\bauthor{\bsnm{{Gibson}}, \binits{S.E.}},
\bauthor{\bsnm{{Foster}}, \binits{D.}},
\bauthor{\bsnm{{Burkepile}}, \binits{J.}},
\bauthor{\bsnm{{de Toma}}, \binits{G.}},
\bauthor{\bsnm{{Stanger}}, \binits{A.}}:
\byear{2006},
\batitle{{The Calm before the Storm: The Link between Quiescent Cavities and
  Coronal Mass Ejections}}.
\bjtitle{\apj}
\bvolume{641},
\bfpage{590}\,--\,\blpage{605}.
doi:\doiurl{10.1086/500446}.
\end{barticle}
\endbibitem

\bibitem[\protect\citeauthoryear{{Gibson}
  \textit{et~al.}}{2010}]{gibson2010-cavities}
\begin{barticle}
\bauthor{\bsnm{{Gibson}}, \binits{S.E.}},
\bauthor{\bsnm{{Kucera}}, \binits{T.A.}},
\bauthor{\bsnm{{Rastawicki}}, \binits{D.}},
\bauthor{\bsnm{{Dove}}, \binits{J.}},
\bauthor{\bsnm{{de Toma}}, \binits{G.}},
\bauthor{\bsnm{{Hao}}, \binits{J.}},
\bauthor{\bsnm{{Hill}}, \binits{S.}},
\bauthor{\bsnm{{Hudson}}, \binits{H.S.}},
\bauthor{\bsnm{{Marqu{\'e}}}, \binits{C.}},
\bauthor{\bsnm{{McIntosh}}, \binits{P.S.}},
\bauthor{\bsnm{{Rachmeler}}, \binits{L.}},
\bauthor{\bsnm{{Reeves}}, \binits{K.K.}},
\bauthor{\bsnm{{Schmieder}}, \binits{B.}},
\bauthor{\bsnm{{Schmit}}, \binits{D.J.}},
\bauthor{\bsnm{{Seaton}}, \binits{D.B.}},
\bauthor{\bsnm{{Sterling}}, \binits{A.C.}},
\bauthor{\bsnm{{Tripathi}}, \binits{D.}},
\bauthor{\bsnm{{Williams}}, \binits{D.R.}},
\bauthor{\bsnm{{Zhang}}, \binits{M.}}:
\byear{2010},
\batitle{{Three-dimensional Morphology of a Coronal Prominence Cavity}}.
\bjtitle{\apj}
\bvolume{724},
\bfpage{1133}\,--\,\blpage{1146}.
doi:\doiurl{10.1088/0004-637X/724/2/1133}.
\end{barticle}
\endbibitem

\bibitem[\protect\citeauthoryear{{Grebinskij}
  \textit{et~al.}}{2000}]{Grebinkij2000}
\begin{barticle}
\bauthor{\bsnm{{Grebinskij}}, \binits{A.}},
\bauthor{\bsnm{{Bogod}}, \binits{V.}},
\bauthor{\bsnm{{Gelfreikh}}, \binits{G.}},
\bauthor{\bsnm{{Urpo}}, \binits{S.}},
\bauthor{\bsnm{{Pohjolainen}}, \binits{S.}},
\bauthor{\bsnm{{Shibasaki}}, \binits{K.}}:
\byear{2000},
\batitle{{Microwave tomography of solar magnetic fields}}.
\bjtitle{\aaps}
\bvolume{144},
\bfpage{169}\,--\,\blpage{180}.
doi:\doiurl{10.1051/aas:2000202}.
\end{barticle}
\endbibitem

\bibitem[\protect\citeauthoryear{{Harvey}}{1969}]{Harvey1969}
\begin{botherref}
\oauthor{\bsnm{{Harvey}}, \binits{J.W.}}:
1969,
{Magnetic Fields Associated with Solar Active-Region Prominences.}
PhD thesis,
University of Colorado at Boulder.
\end{botherref}
\endbibitem

\bibitem[\protect\citeauthoryear{{Heinzel} \textit{et~al.}}{2008}]{Heinzel2008}
\begin{barticle}
\bauthor{\bsnm{{Heinzel}}, \binits{P.}},
\bauthor{\bsnm{{Schmieder}}, \binits{B.}},
\bauthor{\bsnm{{F{\'a}rn{\'{\i}}k}}, \binits{F.}},
\bauthor{\bsnm{{Schwartz}}, \binits{P.}},
\bauthor{\bsnm{{Labrosse}}, \binits{N.}},
\bauthor{\bsnm{{Kotr{\v c}}}, \binits{P.}},
\bauthor{\bsnm{{Anzer}}, \binits{U.}},
\bauthor{\bsnm{{Molodij}}, \binits{G.}},
\bauthor{\bsnm{{Berlicki}}, \binits{A.}},
\bauthor{\bsnm{{DeLuca}}, \binits{E.E.}},
\bauthor{\bsnm{{Golub}}, \binits{L.}},
\bauthor{\bsnm{{Watanabe}}, \binits{T.}},
\bauthor{\bsnm{{Berger}}, \binits{T.}}:
\byear{2008},
\batitle{{Hinode, TRACE, SOHO, and Ground-based Observations of a Quiescent
  Prominence}}.
\bjtitle{\apj}
\bvolume{686},
\bfpage{1383}\,--\,\blpage{1396}.
doi:\doiurl{10.1086/591018}.
\end{barticle}
\endbibitem

\bibitem[\protect\citeauthoryear{{House}}{1977}]{House1977}
\begin{barticle}
\bauthor{\bsnm{{House}}, \binits{L.L.}}:
\byear{1977},
\batitle{{Coronal emission-line polarization from the statistical equilibrium
  of magnetic sublevels. I - Fe XIII}}.
\bjtitle{\apj}
\bvolume{214},
\bfpage{632}\,--\,\blpage{652}.
doi:\doiurl{10.1086/155289}.
\end{barticle}
\endbibitem

\bibitem[\protect\citeauthoryear{{Hudson} \textit{et~al.}}{1999}]{Hudson1999}
\begin{barticle}
\bauthor{\bsnm{{Hudson}}, \binits{H.S.}},
\bauthor{\bsnm{{Acton}}, \binits{L.W.}},
\bauthor{\bsnm{{Harvey}}, \binits{K.L.}},
\bauthor{\bsnm{{McKenzie}}, \binits{D.E.}}:
\byear{1999},
\batitle{{A Stable Filament Cavity with a Hot Core}}.
\bjtitle{\apjl}
\bvolume{513},
\bfpage{L83}\,--\,\blpage{L86}.
doi:\doiurl{10.1086/311892}.
\end{barticle}
\endbibitem

\bibitem[\protect\citeauthoryear{{Jensen}}{2007}]{Jensen_phd2007}
\begin{botherref}
\oauthor{\bsnm{{Jensen}}, \binits{E.A.}}:
2007,
{High frequency Faraday rotation observations of the solar corona}.
PhD thesis,
University of California, Los Angeles.
\end{botherref}
\endbibitem

\bibitem[\protect\citeauthoryear{{Judge}}{2007}]{Judge2007}
\begin{barticle}
\bauthor{\bsnm{{Judge}}, \binits{P.G.}}:
\byear{2007},
\batitle{{Spectral Lines for Polarization Measurements of the Coronal Magnetic
  Field. V. Information Content of Magnetic Dipole Lines}}.
\bjtitle{\apj}
\bvolume{662},
\bfpage{677}\,--\,\blpage{690}.
doi:\doiurl{10.1086/515433}.
\end{barticle}
\endbibitem

\bibitem[\protect\citeauthoryear{{Judge} and {Casini}}{2001}]{Judge2001}
\begin{bchapter}
\bauthor{\bsnm{{Judge}}, \binits{P.G.}},
\bauthor{\bsnm{{Casini}}, \binits{R.}}:
\byear{2001},
\bctitle{{A Synthesis Code for Forbidden Coronal Lines}}.
In: \beditor{\bsnm{{M.~Sigwarth}}} (ed.)
\bbtitle{Advanced Solar Polarimetry -- Theory, Observation, and
  Instrumentation},
\bsertitle{Astron. Soc. Pac.}
\bseriesno{CS-236},
\bfpage{503}.
\end{bchapter}
\endbibitem

\bibitem[\protect\citeauthoryear{{Judge}, {Low}, and
  {Casini}}{2006}]{Judge2006}
\begin{barticle}
\bauthor{\bsnm{{Judge}}, \binits{P.G.}},
\bauthor{\bsnm{{Low}}, \binits{B.C.}},
\bauthor{\bsnm{{Casini}}, \binits{R.}}:
\byear{2006},
\batitle{{Spectral Lines for Polarization Measurements of the Coronal Magnetic
  Field. IV. Stokes Signals in Current-carrying Fields}}.
\bjtitle{\apj}
\bvolume{651},
\bfpage{1229}\,--\,\blpage{1237}.
doi:\doiurl{10.1086/507982}.
\end{barticle}
\endbibitem

\bibitem[\protect\citeauthoryear{{Karpen}, {Antiochos}, and
  {DeVore}}{2012}]{Karpen2012}
\begin{barticle}
\bauthor{\bsnm{{Karpen}}, \binits{J.T.}},
\bauthor{\bsnm{{Antiochos}}, \binits{S.K.}},
\bauthor{\bsnm{{DeVore}}, \binits{C.R.}}:
\byear{2012},
\batitle{{The Mechanisms for the Onset and Explosive Eruption of Coronal Mass
  Ejections and Eruptive Flares}}.
\bjtitle{\apj}
\bvolume{760},
\bfpage{81}.
doi:\doiurl{10.1088/0004-637X/760/1/81}.
\end{barticle}
\endbibitem

\bibitem[\protect\citeauthoryear{{Karpen} \textit{et~al.}}{2001}]{Karpen2001}
\begin{barticle}
\bauthor{\bsnm{{Karpen}}, \binits{J.T.}},
\bauthor{\bsnm{{Antiochos}}, \binits{S.K.}},
\bauthor{\bsnm{{Hohensee}}, \binits{M.}},
\bauthor{\bsnm{{Klimchuk}}, \binits{J.A.}},
\bauthor{\bsnm{{MacNeice}}, \binits{P.J.}}:
\byear{2001},
\batitle{{Are Magnetic Dips Necessary for Prominence Formation?}}
\bjtitle{\apjl}
\bvolume{553},
\bfpage{L85}\,--\,\blpage{L88}.
doi:\doiurl{10.1086/320497}.
\end{barticle}
\endbibitem

\bibitem[\protect\citeauthoryear{{Kramar} and {Inhester}}{2007}]{Kramar2007}
\begin{barticle}
\bauthor{\bsnm{{Kramar}}, \binits{M.}},
\bauthor{\bsnm{{Inhester}}, \binits{B.}}:
\byear{2007},
\batitle{{Inversion of coronal Zeeman and Hanle observations to reconstruct the
  coronal magnetic field}}.
\bjtitle{Mem. Soc. Astron. Ital.}
\bvolume{78},
\bfpage{120}.
\end{barticle}
\endbibitem

\bibitem[\protect\citeauthoryear{{Kramar}, {Inhester}, and
  {Solanki}}{2006}]{Kramar2006}
\begin{barticle}
\bauthor{\bsnm{{Kramar}}, \binits{M.}},
\bauthor{\bsnm{{Inhester}}, \binits{B.}},
\bauthor{\bsnm{{Solanki}}, \binits{S.K.}}:
\byear{2006},
\batitle{{Vector tomography for the coronal magnetic field. I. Longitudinal
  Zeeman effect measurements}}.
\bjtitle{\aap}
\bvolume{456},
\bfpage{665}\,--\,\blpage{673}.
doi:\doiurl{10.1051/0004-6361:20064865}.
\end{barticle}
\endbibitem

\bibitem[\protect\citeauthoryear{{Lin}, {Kuhn}, and {Coulter}}{2004}]{Lin2004}
\begin{barticle}
\bauthor{\bsnm{{Lin}}, \binits{H.}},
\bauthor{\bsnm{{Kuhn}}, \binits{J.R.}},
\bauthor{\bsnm{{Coulter}}, \binits{R.}}:
\byear{2004},
\batitle{{Coronal Magnetic Field Measurements}}.
\bjtitle{\apjl}
\bvolume{613},
\bfpage{L177}\,--\,\blpage{L180}.
doi:\doiurl{10.1086/425217}.
\end{barticle}
\endbibitem

\bibitem[\protect\citeauthoryear{{Lin}, {Penn}, and {Tomczyk}}{2000}]{Lin2000}
\begin{barticle}
\bauthor{\bsnm{{Lin}}, \binits{H.}},
\bauthor{\bsnm{{Penn}}, \binits{M.J.}},
\bauthor{\bsnm{{Tomczyk}}, \binits{S.}}:
\byear{2000},
\batitle{{A New Precise Measurement of the Coronal Magnetic Field Strength}}.
\bjtitle{\apjl}
\bvolume{541},
\bfpage{L83}\,--\,\blpage{L86}.
doi:\doiurl{10.1086/312900}.
\end{barticle}
\endbibitem

\bibitem[\protect\citeauthoryear{{Liu} and {Lin}}{2008}]{Liu2008}
\begin{barticle}
\bauthor{\bsnm{{Liu}}, \binits{Y.}},
\bauthor{\bsnm{{Lin}}, \binits{H.}}:
\byear{2008},
\batitle{{Observational Test of Coronal Magnetic Field Models. I. Comparison
  with Potential Field Model}}.
\bjtitle{\apj}
\bvolume{680},
\bfpage{1496}\,--\,\blpage{1507}.
doi:\doiurl{10.1086/588645}.
\end{barticle}
\endbibitem

\bibitem[\protect\citeauthoryear{{Low} and {Hundhausen}}{1995}]{LowHund1995}
\begin{barticle}
\bauthor{\bsnm{{Low}}, \binits{B.C.}},
\bauthor{\bsnm{{Hundhausen}}, \binits{J.R.}}:
\byear{1995},
\batitle{{Magnetostatic structures of the solar corona. 2: The magnetic
  topology of quiescent prominences}}.
\bjtitle{\apj}
\bvolume{443},
\bfpage{818}\,--\,\blpage{836}.
doi:\doiurl{10.1086/175572}.
\end{barticle}
\endbibitem

\bibitem[\protect\citeauthoryear{{Luna}, {Karpen}, and
  {DeVore}}{2012}]{Luna2012}
\begin{barticle}
\bauthor{\bsnm{{Luna}}, \binits{M.}},
\bauthor{\bsnm{{Karpen}}, \binits{J.T.}},
\bauthor{\bsnm{{DeVore}}, \binits{C.R.}}:
\byear{2012},
\batitle{{Formation and Evolution of a Multi-threaded Solar Prominence}}.
\bjtitle{\apj}
\bvolume{746},
\bfpage{30}.
doi:\doiurl{10.1088/0004-637X/746/1/30}.
\end{barticle}
\endbibitem

\bibitem[\protect\citeauthoryear{{Mackay} \textit{et~al.}}{2010}]{Mackay2010}
\begin{barticle}
\bauthor{\bsnm{{Mackay}}, \binits{D.H.}},
\bauthor{\bsnm{{Karpen}}, \binits{J.T.}},
\bauthor{\bsnm{{Ballester}}, \binits{J.L.}},
\bauthor{\bsnm{{Schmieder}}, \binits{B.}},
\bauthor{\bsnm{{Aulanier}}, \binits{G.}}:
\byear{2010},
\batitle{{Physics of Solar Prominences: II-Magnetic Structure and Dynamics}}.
\bjtitle{\ssr}
\bvolume{151},
\bfpage{333}\,--\,\blpage{399}.
doi:\doiurl{10.1007/s11214-010-9628-0}.
\end{barticle}
\endbibitem

\bibitem[\protect\citeauthoryear{{Mari{\v c}i{\'c}}
  \textit{et~al.}}{2004}]{Maricic2004}
\begin{barticle}
\bauthor{\bsnm{{Mari{\v c}i{\'c}}}, \binits{D.}},
\bauthor{\bsnm{{Vr{\v s}nak}}, \binits{B.}},
\bauthor{\bsnm{{Stanger}}, \binits{A.L.}},
\bauthor{\bsnm{{Veronig}}, \binits{A.}}:
\byear{2004},
\batitle{{Coronal Mass Ejection of 15 May 2001: I. Evolution of Morphological
  Features of the Eruption}}.
\bjtitle{\solphys}
\bvolume{225},
\bfpage{337}\,--\,\blpage{353}.
doi:\doiurl{10.1007/s11207-004-3748-1}.
\end{barticle}
\endbibitem

\bibitem[\protect\citeauthoryear{{Patzold} \textit{et~al.}}{1987}]{Patzold1987}
\begin{barticle}
\bauthor{\bsnm{{Patzold}}, \binits{M.}},
\bauthor{\bsnm{{Bird}}, \binits{M.K.}},
\bauthor{\bsnm{{Volland}}, \binits{H.}},
\bauthor{\bsnm{{Levy}}, \binits{G.S.}},
\bauthor{\bsnm{{Seidel}}, \binits{B.L.}},
\bauthor{\bsnm{{Stelzried}}, \binits{C.T.}}:
\byear{1987},
\batitle{{The mean coronal magnetic field determined from HELIOS Faraday
  rotation measurements}}.
\bjtitle{\solphys}
\bvolume{109},
\bfpage{91}\,--\,\blpage{105}.
doi:\doiurl{10.1007/BF00167401}.
\end{barticle}
\endbibitem

\bibitem[\protect\citeauthoryear{{Querfeld}}{1977}]{Querfeld1977}
\begin{bchapter}
\bauthor{\bsnm{{Querfeld}}, \binits{C.W.}}:
\byear{1977},
\bctitle{{A near-infrared coronal emission-line polarimeter}}.
In: \beditor{\bsnm{{Azzam}}, \binits{R.M.A.}},
\beditor{\bsnm{{Coffeen}}, \binits{D.L.}} (eds.)
\bbtitle{Optical Polarimetry: Instrumentation and Applications},
\bsertitle{Proc. SPIE 0112}
\bseriesno{112},
\bfpage{200}\,--\,\blpage{208}.
\end{bchapter}
\endbibitem

\bibitem[\protect\citeauthoryear{{Rachmeler}, {Casini}, and
  {Gibson}}{2012}]{Rachmeler2012}
\begin{bchapter}
\bauthor{\bsnm{{Rachmeler}}, \binits{L.A.}},
\bauthor{\bsnm{{Casini}}, \binits{R.}},
\bauthor{\bsnm{{Gibson}}, \binits{S.E.}}:
\byear{2012},
\bctitle{{Interpreting Coronal Polarization Observations}}.
In: \beditor{\bsnm{{Rimmele}}, \binits{T.R.}},
\beditor{\bsnm{{Tritschler}}, \binits{A.}},
\beditor{\bsnm{{W{\"o}ger}}, \binits{F.}},
\beditor{\bsnm{{Collados Vera}}, \binits{M.}},
\beditor{\bsnm{{Socas-Navarro}}, \binits{H.}},
\beditor{\bsnm{{Schlichenmaier}}, \binits{R.}},
\beditor{\bsnm{{Carlsson}}, \binits{M.}},
\beditor{\bsnm{{Berger}}, \binits{T.}},
\beditor{\bsnm{{Cadavid}}, \binits{A.}},
\beditor{\bsnm{{Gilbert}}, \binits{P.R.}},
\beditor{\bsnm{{Goode}}, \binits{P.R.}},
\beditor{\bsnm{{Kn{\"o}lker}}, \binits{M.}} (eds.)
\bbtitle{Second ATST-EAST Meeting: Magnetic Fields from the Photosphere to the
  Corona.},
\bsertitle{Astron. Soc. Pac.}
\bseriesno{CS-463},
\bfpage{227}.
\end{bchapter}
\endbibitem

\bibitem[\protect\citeauthoryear{{Reeves} \textit{et~al.}}{2012}]{Reeves2012}
\begin{barticle}
\bauthor{\bsnm{{Reeves}}, \binits{K.K.}},
\bauthor{\bsnm{{Gibson}}, \binits{S.E.}},
\bauthor{\bsnm{{Kucera}}, \binits{T.A.}},
\bauthor{\bsnm{{Hudson}}, \binits{H.S.}},
\bauthor{\bsnm{{Kano}}, \binits{R.}}:
\byear{2012},
\batitle{{Thermal Properties of a Solar Coronal Cavity Observed with the X-Ray
  Telescope on Hinode}}.
\bjtitle{\apj}
\bvolume{746},
\bfpage{146}.
doi:\doiurl{10.1088/0004-637X/746/2/146}.
\end{barticle}
\endbibitem

\bibitem[\protect\citeauthoryear{{R{\'e}gnier}, {Walsh}, and
  {Alexander}}{2011}]{Regnier2011}
\begin{barticle}
\bauthor{\bsnm{{R{\'e}gnier}}, \binits{S.}},
\bauthor{\bsnm{{Walsh}}, \binits{R.W.}},
\bauthor{\bsnm{{Alexander}}, \binits{C.E.}}:
\byear{2011},
\batitle{{A new look at a polar crown cavity as observed by SDO/AIA. Structure
  and dynamics}}.
\bjtitle{\aap}
\bvolume{533},
\bfpage{L1}.
doi:\doiurl{10.1051/0004-6361/201117381}.
\end{barticle}
\endbibitem

\bibitem[\protect\citeauthoryear{{Schmit} and {Gibson}}{2011}]{Schmit2011}
\begin{barticle}
\bauthor{\bsnm{{Schmit}}, \binits{D.J.}},
\bauthor{\bsnm{{Gibson}}, \binits{S.E.}}:
\byear{2011},
\batitle{{Forward Modeling Cavity Density: A Multi-instrument Diagnostic}}.
\bjtitle{\apj}
\bvolume{733},
\bfpage{1}.
doi:\doiurl{10.1088/0004-637X/733/1/1}.
\end{barticle}
\endbibitem

\bibitem[\protect\citeauthoryear{{Tomczyk} \textit{et~al.}}{2008}]{Tomczyk2008}
\begin{barticle}
\bauthor{\bsnm{{Tomczyk}}, \binits{S.}},
\bauthor{\bsnm{{Card}}, \binits{G.L.}},
\bauthor{\bsnm{{Darnell}}, \binits{T.}},
\bauthor{\bsnm{{Elmore}}, \binits{D.F.}},
\bauthor{\bsnm{{Lull}}, \binits{R.}},
\bauthor{\bsnm{{Nelson}}, \binits{P.G.}},
\bauthor{\bsnm{{Streander}}, \binits{K.V.}},
\bauthor{\bsnm{{Burkepile}}, \binits{J.}},
\bauthor{\bsnm{{Casini}}, \binits{R.}},
\bauthor{\bsnm{{Judge}}, \binits{P.G.}}:
\byear{2008},
\batitle{{An Instrument to Measure Coronal Emission Line Polarization}}.
\bjtitle{\solphys}
\bvolume{247},
\bfpage{411}\,--\,\blpage{428}.
doi:\doiurl{10.1007/s11207-007-9103-6}.
\end{barticle}
\endbibitem

\bibitem[\protect\citeauthoryear{{Trujillo Bueno}}{2001}]{Trujillo2001}
\begin{bchapter}
\bauthor{\bsnm{{Trujillo Bueno}}, \binits{J.}}:
\byear{2001},
\bctitle{{Atomic Polarization and the Hanle Effect}}.
In: \beditor{\bsnm{{Sigwarth}}, \binits{M.}} (ed.)
\bbtitle{Advanced Solar Polarimetry -- Theory, Observation, and
  Instrumentation},
\bsertitle{Astron. Soc. Pac.}
\bseriesno{CS-236},
\bfpage{161}.
\end{bchapter}
\endbibitem

\bibitem[\protect\citeauthoryear{{van Vleck}}{1925}]{vanVleck1925}
\begin{barticle}
\bauthor{\bsnm{{van Vleck}}, \binits{J.H.}}:
\byear{1925},
\batitle{{On the Quantum Theory of the Polarization of Resonance Radiation in
  Magnetic Fields}}.
\bjtitle{Proc. Nat. Acad. Sci.}
\bvolume{11},
\bfpage{612}\,--\,\blpage{618}.
doi:\doiurl{10.1073/pnas.11.10.612}.
\end{barticle}
\endbibitem

\bibitem[\protect\citeauthoryear{{White} and {Kundu}}{1997}]{White1997}
\begin{barticle}
\bauthor{\bsnm{{White}}, \binits{S.M.}},
\bauthor{\bsnm{{Kundu}}, \binits{M.R.}}:
\byear{1997},
\batitle{{Radio Observations of Gyroresonance Emission from Coronal Magnetic
  Fields}}.
\bjtitle{\solphys}
\bvolume{174},
\bfpage{31}\,--\,\blpage{52}.
doi:\doiurl{10.1023/A:1004975528106}.
\end{barticle}
\endbibitem

\end{thebibliography}

\end{article} 

\end{document}